\documentclass[12pt]{article}
\usepackage{a4,graphics,latexsym}
\usepackage{subfigure}

\def\bn{\mbox{\boldmath $N$}}
\def\bp{\mbox{\boldmath $\phi$}}
\def\ba{\mbox{\boldmath $a$}}
\def\bx{\mbox{\boldmath $X$}}

\input epsf
\begin{document}

\newcommand{\ind}{\hspace{0.3in}}
\newcommand\lsim{\mathrel{\rlap{\lower4pt\hbox{\hskip1pt$\sim$}}
    \raise1pt\hbox{$<$}}}
\newcommand\gsim{\mathrel{\rlap{\lower4pt\hbox{\hskip1pt$\sim$}}
    \raise1pt\hbox{$>$}}}

\begin{flushright}
SHEP 99/03\\
hep-ph/9904315\\
\end{flushright}

\vspace{.4in} 

\begin{center}
\Large\textbf {Preheating in Supersymmetric Hybrid Inflation}

\vspace{0.2in}

\normalsize
M. Bastero-Gil, S. F. King and J. Sanderson

\emph{Department of Physics and Astronomy, University of Southampton,\\
Southampton, SO17 1BJ, U.K.}

\end{center}

\vspace{0.1in}

\footnotesize

\begin{center}
\emph{ABSTRACT}
\end{center}

We study preheating in a general class of supersymmetric hybrid
inflation model. Supersymmetry leads to only one coupling constant in
the potential and thus only one natural frequency of oscillation for
the homogeneous fields, whose classical evolution consequently differs
from that of a general (non-supersymmetric) hybrid model. We
emphasise the importance of mixing effects in these models which can
significantly change the rate of production of particles.
We perform a general study of the rate of production of the particles
associated with the homogeneous fields, and show how preheating is
efficient in producing these quanta.  Preheating of other particle
species will be model dependent, and in order to investigate this we
consider a realistic working model of supersymmetric hybrid inflation
which solves the strong-CP problem via an approximate Peccei-Quinn
symmetry, which was proposed by us previously.  We study axion
production in this model and show that properly taking into account
the mixing between the fields suppresses the axion production, yet
enhances the production of other particles. Finally we demonstrate the
importance of backreaction effects in this model which have the effect
of shutting off axion production, leaving the axion safely within
experimental bounds.

\vspace{.5in}

\normalsize 

\newpage

\section{Introduction}

Reheating in the post-inflationary Universe is a very important
process, since it describes the formation of all matter and energy in
the Universe today \cite{ref1}. However, it has only relatively
recently \cite{ref2} been realised that in certain cases the process
of reheating involves a stage of explosive production of particles by
parametric resonance which cannot be described by the usual methods of
the elementary reheating theory \cite{ref8}. This stage of
`preheating' is a non-perturbative and out of equilibrium process and as such is
a very involved and little understood proceeding. In fact there have
been complex analytical methods of preheating developed for only the
very simplest of single-field inflationary potentials \cite{ref7}. It
is generally numerically easy to obtain results for the first stages
of particle production, yet a complete, self-consistent analysis
should include all effects of backreaction of produced particles and
their rescattering. At present lattice simulations (such as in
\cite{ref31,ref3}) must be performed to fully study these effects.
Nevertheless, much progress is being made in developing methods to
cope with the rich topic of preheating \cite{ref7}.

For a while now, the advantages of using more than one scalar field to
implement inflation have been well appreciated. Hybrid inflation
models \cite{H} are probably the most popular of all present
inflationary models. They succeed in removing the need for fine-tuning
of the inflaton potential by providing a natural means to end
inflation. Most importantly however, they allow inflation to occur at
a relatively low energy density using small scale fields avoiding
strong CMB constraints, a task which is virtually impossible to
achieve using only one field. Thus, hybrid-type models are attractive
from the point of view of particle physics and in particular are
naturally well suited to realisations in supersymmetry (SUSY) where
there is an abundance of scalar fields and flat directions.

Preheating in hybrid inflation has been studied \cite{ref4} using the
standard toy hybrid potential of two fields $V(\phi_1,\phi_2)$,
coupled to an extra, massless scalar field, $\chi$ by the interaction 
$\frac{1}{2}(h_1^2\phi_1^2+h_2^2 \phi_2^2)\chi^2$. It is found
\cite{ref4} that over most of the parameter space, preheating is
rather inefficient in hybrid inflation with little production of
$\phi_1,\phi_2$ particles although production of $\chi$ particles is
possible for certain values of $h_1, h_2$. However, the results
concerning the production of $\phi_1$ and $\phi_2$ particles may
change in a supersymmetric version of the model, where the potential
is derived from a supersymmetric superpotential of the kind $W \sim
\kappa \phi_1 \phi_2^2+\cdots$, such that the same coupling will
give the quartic self-interaction for the $\phi_2$ field and the
interaction term between $\phi_2$ and the inflaton
$\phi_1$. Therefore, supersymmetry will ensure that the masses of both
fields at the global minimum of the potential are not only the same order
of magnitude but $equal$. This in turn will give rise to a
non-chaotic  evolution of the classical homogeneous fields at the
beginning of the preheating/reheating era. The classical fields will
begin to oscillate with relatively large, little damped amplitudes,  which will
favour production of particles occurring. 

In addition, in hybrid models inflation ends with a phase transition
when the inflaton field reaches a critical value.  At this point, the
second field $\phi_2$ is sited at a maximum of the potential with
vanishing mass. Once the inflaton passes this point, the mass of
$\phi_2$ becomes negative allowing both fields to roll down the
potential towards the global minimum. It is clear that across the
region of the spinodal instability (negative curvature)
\cite{refDan1}, production of $\phi_2$ particles can be quite
intensive. Therefore, previous to any oscillation of the homogeneous
fields, preheating will begin with an initial burst of $\phi_2$
production.  If the amplitude of the oscillations is large enough,
this will also happen in the successive oscillations every time the
classical fields enter the region of the spinodal instability.
Although this only directly affects the rate of production of $\phi_2$
particles, the effect is propagated into the production of
$\phi_1$ particles through the mixing term in the mass squared matrix.

The layout of the paper is as follows: section 2 briefly describes the theory 
of preheating as applied to many scalar fields and we point out that
the coupling between fields should in general be taken into account;
in section 3 we study a general model of supersymmetric hybrid
inflation, focusing on the issue of production of $\phi_1$ and $\phi_2$
particles; in order to motivate production of other particles, which
is model dependent, in  
section 4 we review a particular model of inflation where the strong-CP
problem is solved via a Peccei-Quinn symmetry which leads to the
production of axions. It will be important to calculate the
effects of preheating in this model since the number of axions produced during
reheating must be tightly constrained in order to be cosmologically
acceptable. The results of the numerical study of the
first stages of preheating in this model are presented, and compared
to those results without including mixing terms. The first effects of
backreaction of produced particles is studied and the total number of
each type of particle calculated.

\section{Theory of preheating with multiple fields}

We begin with a model of the early Universe described by a potential,
$V(\phi_\alpha)$, in an expanding Universe with scale factor
$a(t)$. Amongst the fields $\phi_\alpha$ in the potential, there are
degrees of freedom corresponding to scalar fields in which all energy
is concentrated during inflation. In the usually considered
single-field inflation this is the slowly rolling inflaton, $\phi$,
with effective potential $V(\phi)$. In the hybrid scenario there are
two such fields important to inflation, one in which the potential is
relatively flat during inflation which acts as the inflaton field,
$\phi_1$, and one which sits in a false vacuum, $\phi_2$, and causes
the end of inflation in the phase transition to its true vacuum. 
All other fields are not important to inflation serving only perhaps
to provide the constant vacuum energy during inflation.

In order to investigate the reheating of the Universe and determine
the extent of the effects of preheating, we must first study the
evolution of these fields as inflation ends and beyond into the regime
of coherent oscillations.

We may write the full quantum fields as:
\begin{equation}
\bp_{\alpha}(\textbf{x},t)=\phi_{\alpha}(t)
+\delta\phi_{\alpha}(\textbf{x},t), \label{phidelta} 
\end{equation}
where $\delta\phi_{\alpha}(\textbf{x},t)$ represents the quantum fluctuations
about the homogeneous zero modes, $\phi_{\alpha}(t)$, and necessarily
satisfies $\langle\delta\phi_{\alpha}(\textbf{x},t)\rangle=0$.

Neglecting ensuing effects of particle production, we proceed by
considering the motion classically, solving the equations of
motion for the classical homogeneous fields $\phi_{\alpha}(t)$, but using
quantum fluctuations $\sim H_0/2\pi$ (where $H_0$ is the Hubble
constant during inflation) in the initial conditions at
the end of inflation to initiate the spontaneous symmetry breaking.

The equations of motion governing the behaviour of the background
fields, $\phi_{\alpha}(t)$, are:
\begin{equation}
\ddot{\phi}_{\alpha}+3H\dot{\phi_{\alpha}}+\partial V/\partial\phi_{\alpha}=0,     \label{cleom}
\end{equation}

where the Hubble constant is given by the Friedmann equation,
\begin{equation}
H^2(t) =
\frac{8\pi}{3M_P^2}\left[
\frac{1}{2}\sum_{\alpha}\dot{\phi_{\alpha}}^2+V\right], 
\label{H} 
\end{equation}
and $H=\dot{a}/a$. 

Here, the subscript $\alpha$ runs over all the classical scalar fields in
the theory relevant to inflation and $V=V(\phi_{\alpha})$ is the effective
potential of these fields. During inflation all energy is contained in
this potential and the Universe is in a vacuum-like state with
vanishing temperature, entropy and particle number densities. The
inflaton field, $\phi$, is slowly rolling with the friction term
dominating in Eq. (\ref{cleom}) so that:
\begin{equation}
\dot{\phi}\simeq\frac{-1}{3H_0}\frac{\partial
V}{\partial\phi}. \label{slowroll}  
\end{equation}
This also describes the relevant situation for each classical field at the
beginning of the reheating era.

The equations of motion for the quantum
fluctuations of the scalar fields in linear perturbation theory are
given by:
\begin{equation}
\ddot{\delta\phi_k}_\alpha +
3H\dot{\delta\phi_k}_\alpha+\frac{k^2}{a^2}{\delta\phi_k}_\alpha +
\mathcal{M}^2_{\alpha \beta}{\delta\phi_k}_\beta = 0,
\label{queom}
\end{equation}
where ${\delta\phi_k}_\alpha(t)$ are the time dependent Fourier modes (in
comoving momentum space $\textbf{k}$) of the quantum fluctuations
$\delta\phi_\alpha$,
\begin{equation}
\delta\phi_\alpha(\textbf{x},t)=\int\frac{d^3k}{\sqrt{2}(2\pi)^{3/2}}
\left[\hat{a}_k{\delta\phi_k}_\alpha(t)e^{-i\textbf{k.x}}
+\hat{a}^\dagger_k{\delta\phi_k}_\alpha^*(t)e^{i\textbf{k.x}}\right], 
\end{equation}
and $\hat{a}^\dagger_k$ and $\hat{a}_k$ are creation and annihilation
operators.
Thus, Eq. (\ref{queom}) describes the interactions between the
classical fields, $\phi_{\alpha}$, and the quantum fluctuations,
$\delta\phi_\alpha$. When neglecting backreaction effects due to the
particles produced, the Hubble constant, $H(t)$, and the mass squared
matrix,
$\mathcal{M}^2_{\alpha\beta}(t)=
\partial^2V/\partial\phi_\alpha\partial\phi_\beta$, 
are functions of the classical fields only, as given by the classical
equations of motion  (\ref{cleom}) and (\ref{H}). It should be clear that
 the equations of motion for 
the quantum fluctuations will in general be coupled through mixing
terms in the mass matrix $\mathcal{M}^2_{\alpha\beta}\not= 0$ (for
$\alpha\not=\beta$); in other words the rate of production of each
separate particle is dependent on the rate of production of the
others, {\it{even}} at the initial stage of preheating when
backreaction effects are not yet important.

Changing to comoving fields,
${\psi_k}_\alpha=a^{3/2}{{\delta\phi_k}_\alpha}$, these equations can
be simplified further to:
\begin{equation}
\ddot{\psi_k}_\alpha +
\left(\frac{k^2}{a^2}-\epsilon\right){\psi_k}_\alpha +
\mathcal{M}^2_{\alpha\beta}{\psi_k}_\beta = 0,	
							\label{psieq}
\end{equation}
where one usually neglects the pressure term, $\epsilon =
\frac{3}{4}(H^2+2\ddot{a}/a){\psi_k}_\alpha$ which is a small correction
that approaches zero as the Universe approaches matter domination
$(a\sim t^{2/3})$. Typically the transition between the vacuum
dominated state of inflation and that of matter domination during
reheating is fairly rapid, so this term will be negligible.

The solutions of (\ref{psieq}), ${\psi_k}_\alpha(t)$, are complex functions
whose initial values at the end of inflation at time $t_0$ are taken
to be \cite{refDan2}:
\begin{equation}
{\psi_k}_\alpha(t_0)=1/\sqrt{2{\omega_k}_\alpha(t_0)},\;\;\;
\dot{{\psi}_k}_\alpha(t_0)=-i\sqrt{{\omega_k}_\alpha(t_0)/2}, 
\end{equation}
where,
\begin{equation}
{\omega_k}_\alpha^2 = k^2/a^2 + \mathcal{M}^2_{\alpha\alpha}.
\end{equation}
The comoving occupation number of particles of wave number $\textbf{k}$ is
an adiabatic invariant defined by:
\begin{equation}
{n_k}_\alpha=\frac{{\omega_k}_\alpha}{2}\left(\frac{|\dot{{\psi}_k}_\alpha|^2}{{\omega_k}_\alpha^2}+|{{\psi}_k}_\alpha|^2\right)-\frac{1}{2}.								\label{occno}
\end{equation}
An instability in the growth of a given mode, ${{\psi}_k}_\alpha$,
corresponds to an exponential increase in the occupation number of
particles, ${n_k}_\alpha\propto
\exp{(2{\mu_k}_\alpha\bar{m}t)}$. $\bar{m}$ denotes the characteristic
frequency of oscillations of the classical fields, given by their mass at
the global minimum. The efficiency of preheating of a given particle
with a given mode $\textbf{k}$ is measured by the growth parameter
${\mu_k}_\alpha$.

Of course to be fully self-consistent, the potential and its
derivatives in the equations for the classical fields and Hubble
constant, (\ref{cleom}), (\ref{H}), should contain the full 
quantum fields in the  
theory, $V(\bp_\alpha)$, taking into account the extra particles
being produced. In the Hartree approximation this is done via
$\langle\delta\phi_\alpha^2\rangle\ne 0$, where  
the expectation value contains all modes:
\begin{equation}
\langle\delta\phi_\alpha^2\rangle =
\int\frac{k^2dk}{2\pi^2}|{\delta\phi_k}_\alpha|^2. 
\end{equation}
The backreaction on the classical fields will not be important in the
first stage of the evolution when $\langle\delta\phi_\alpha^2\rangle\simeq
0$, but the modes will grow and eventually dominate over the other
terms. The classical evolution will be altered and in turn this will
affect the quantum equations (\ref{queom}) which also must include
further terms corresponding to produced particles. 

As a first step we may neglect production of particles altogether
and study the evolution classically. Once  the behaviour of the
classical fields in a specific model is given, we may try to rewrite the
evolution equation for each quantum fluctuation, Eq. (\ref{psieq}), in
the form of a Mathieu equation \cite{mathieu},
\begin{equation}
{\psi_k}^{''}_\alpha +
(A_{\alpha}(k)-2q_{\alpha}\cos{2z}){\psi_k}_\alpha = 0,\label{mat} 
\end{equation}
where dash denotes differentiation with respect to $z$, and $z\propto
\bar{m} t$. The $(A,q)$ plane is divided into instability/stability
regions, such that in the unstable regions the solution grows
exponentially, $\psi_k \propto e^{\mu_k z}$, and this can be
interpreted as production of particles for the associated mode.
Therefore, once the parameters $A_{\alpha}(k)$ and $q_\alpha$ are
known for a particular model we can get some insight about which
particles are more likely to be produced, without the need of
numerical integration.

In the next section  we  will apply the above receipt to a general
supersymmetric hybrid model. First we analyse the motion
of the classical fields when they begin to oscillate, focusing
afterwards on the implications this will have for the production of their
own quantum fluctuations during preheating.

\section{Supersymmetric hybrid inflation.} 

In the hybrid scenario \cite{H} with just two real fields, 
the potential is given by,
\begin{equation}
V(\phi, N)= \frac{\lambda}{4} \left( N^2 -
2 \Lambda^2\right)^2 + \frac{g^2}{2} \phi^2 N^2 + \frac{1}{2} m_\phi^2
\phi^2 \,,
\label{Vhyb}
\end{equation}
where $\phi$ will play the role of the inflaton. 
During inflation, the field $N$ is trapped in its false vacuum, $N=0$,
whilst the $\phi$ field slowly rolls along the potential from above
the critical value $\phi_c= (\sqrt{2 \lambda}/g ) \Lambda$ towards the
origin. Once $\phi$ 
reaches the critical value, the $N$ mass changes sign and becomes
negative, allowing $N$ to roll away from its zero value towards the
global minimum, which in the positive quadrant is given by,
\begin{equation}
\phi_0=0 \,,\;\;\;\; N_0= \sqrt{2} \Lambda \,.
\end{equation}
The fields then oscillate about their true vacuum values ending
inflation and beginning the preheating/reheating phase. 

Integrating the classical equations of motion, we obtain
the trajectory in the $\phi-N$ plane. In many cases after the
initial symmetry breaking, one field may oscillate with an amplitude
much larger than that of the other field which is essentially fixed at
its VEV. In this case almost all the energy is contained in just one
field and reheating is then similar to the single field case. This
will greatly simplify the investigation of preheating, although in
general it will not be possible. Alternatively both fields may
oscillate, and the motion may be chaotic. Having one or another
situation will depend  on the ratio of the coupling constants, 
$\lambda/g^2$, and they  have been
analysed in \cite{ref4}, the chaotic motion being typical of a
potential with $\lambda \approx g^2$. However, a
different scenario 
will arise in the supersymmetric hybrid models of inflation for
which the exact relation $\lambda= g^2/2$ holds. 

In supersymmetric hybrid inflation, the potential in Eq. (\ref{Vhyb})
can be  derived from the superpotential, 
\begin{equation}
W= \kappa \Phi (\rm{N}^2 - \Lambda^2)  \label{supot}\,,
\end{equation}
where $\Phi$ and $\rm{N}$ are complex superfields, $\phi$ and $N$ in
the potential being the real components of the associated canonically
normalised complex fields, and $\kappa^2=\lambda= g^2/2$. In addition,
the inflaton mass $m_\phi$ will be given either by a soft SUSY-breaking
mass or else generated by radiative corrections to the inflaton
potential. In either case, it has to be smaller than the combination
$\kappa \Lambda$ to ensure a potential flat enough during inflation
to allow for the slow rolling scenario \cite{liddle}, that is,
\begin{eqnarray}
\epsilon &=& \frac{M_P^2}{16 \pi} \left(\frac{V'}{V}\right)^2=
\frac{1}{16 \pi}\left( \frac{M_P \phi}{\Lambda^2} \right)^2 
\left(\frac{m_\phi}{ \kappa \Lambda} \right)^4   \ll 1\,,
\label{epsilon} \\ 
|\eta| &=& \frac{M_P^2}{8 \pi} \left|\frac{V''}{V}\right| =
\frac{1}{8 \pi} \left(\frac{M_P}{\Lambda}\right)^2 
\left(\frac{m_\phi}{ \kappa \Lambda} \right)^2 \ll 1 \,,
\label{eta}
\end{eqnarray}
where prime denotes derivative with respect to $\phi$.
Moreover, the model should be able to
reproduce the correct level of density perturbation, responsible for
the large scale structure in the Universe, accordingly to the COBE
anisotropy measurements. The spectrum of the density perturbations is
given by the quantity \cite{liddle},
\begin{equation}
\delta^2_H= \frac{32}{75} \frac{V}{M^4_P} \frac{1}{\epsilon_*} \,,
\label{cobe}
\end{equation}
where 
$\epsilon_*$  is defined  roughly 60 e-folds before the end of inflation,
and  the COBE value is $\delta_H=1.95 \times 10^{-5}$ \cite{COBE}.
Assuming  $|\eta| < 1/60 $, then $\phi_*= \phi_c e^{ 60 \eta}
\approx \phi_c$, and Eq. (\ref{cobe}) yields the constraint, 
\begin{equation} 
\kappa \Lambda \simeq 2.36 \times 10^{10} \left(\kappa
m_{\phi}\right)^{2/5}\, GeV\,,  
\label{cobecon}
\end{equation}
or similarly,
\begin{equation}
\kappa \Lambda \simeq 1.27 \times 10^{15} |\eta| \,.
\label{cobecon1}
\end{equation}
The value $|\eta| \simeq 0.01$ is also a reasonable assumption in
order not to generate too large a tilt in the spectral index, $n=1 + 2
\eta - 6 \epsilon$, whose present observational constraint \cite{sindex} is
$|n-1|< 0.2$. 

The superpotential in Eq. (\ref{supot}) provides not only the
couplings of the inflaton and hybrid field $N$ neccessary to achieve
the hybrid mechanism, but also a non-vanishing constant vacuum energy
$V(0)$ through the scale $\Lambda$, which is also related to the
critical value $\phi_c$. Hybrid inflation models derived from this
kind of superpotential are called F-term hybrid inflation.  A
different version of SUSY hybrid inflation is provided by D-term
inflation \cite{DI}, where the critical value and the constant
potential energy are related to the Fayet-Illiopoulus term coming from
an anomalous $U(1)$ symmetry. Alternatively the constant potential
$V(0)$ may owe its origin to the hidden sector of a supergravity
(SUGRA) theory \cite{SUGRA}. Nevertheless, the ultimate origin of
$V(0)$ and the scale $\phi_c$ is not relevant for the discussion of
preheating. We only demand a common coupling giving the quartic
interaction term for $N$ and the interaction between $\phi$ and $N$,
and the condition that the potential vanishes at the global
minimum. Therefore, the SUSY hybrid potential can be written as:
\begin{equation}
V_0(\phi,N) = \frac{1}{4}{\kappa}^2 (N^2 -2 \phi^2_c)^2 + {\kappa}^2
N^2 \phi^2 + \frac{1}{2} m_\phi^2 \phi^2. 
\end{equation}
where the origin of $\phi_c$ is left unspecified. The slow-rolling and
COBE constraints, Eq.s (\ref{epsilon}-\ref{cobecon1}), remain the same with
the replacement $\phi_c=\Lambda$.

\subsection{Classical field evolution \label{clshyb}}

Whatever the values of the parameters $\kappa$, $\phi_c$ and
$m_\phi$ which satisfy the above constraints, the behaviour of the
classical fields when they begin to oscillate will follow a
common pattern, dictated  by the form of the potential, 
 and we can make some general observations which  are
independent of the particular values of the parameters chosen:

\noindent \emph{i)} The maximum attainable amplitude of the
oscillations of the $\phi$ field, $\phi_{c}=N_0/\sqrt{2}$, is
comparable to (more precisely is a 
factor of $\sqrt{2}$ smaller than) that of the $N$ field, $N_0$.

\noindent \emph{ii)} There is only $one$ natural frequency of
oscillation, given by the masses of the fields at the global minimum,
\begin{equation}
\bar{m}_\phi = \bar{m}_N= \sqrt{2} \kappa N_0 \,.
\end{equation}
This is quite different to the 
situation in most of the parameter regime of the standard
(non-supersymmetric) hybrid potential, and it is due to the exact
relation $\lambda= g^2/2$ imposed by SUSY. 

\noindent \emph{iii)}
In the limit $m_\phi \ll \kappa \phi_c$, and due to the fact that
both fields have the same mass at the global minimum, there is a
particular solution of the classical equations such that,
\begin{equation}
\frac{\partial V_0}{\partial \phi} \propto \frac{\partial
V_0}{\partial N} \,,
\end{equation}
that is, the fields' trajectory towards the minimum of the potential
describes a straight line:
\begin{eqnarray}
N=\pm\sqrt{2}(\phi_{c}-\phi),
\label{traj}
\end{eqnarray}
in the $\phi-N$ plane. That the fields follow such a trajectory or not
will depend on the initial conditions, otherwise fixed by the
inflationary dynamics and given by:
\begin{eqnarray}
& &\phi_{in} = \phi_{c}\pm \frac{H_0}{2\pi},\hspace{0.3in}
N_{in} = \frac{H_0}{2\pi}, \nonumber \\ 
& &\dot{\phi}_{in} =\frac{-1}{3H_0}\frac{\partial
V_0}{\partial\phi}(\phi_{in},N_{in}),\hspace{0.2in}
\dot{N}_{in} = \frac{-1}{3H_0}\frac{\partial V_0}{\partial N}(\phi_{in},N_{in}). \nonumber
\end{eqnarray}
Since the initial velocities, $\dot{\phi}_{initial}$ and
$\dot{N}_{initial}$, are very small slow roll values, the fields will tend to
move following the gradient field of the potential towards the
minimum, and will end in the straight line trajectory almost
exactly. In effect, we will have one oscillating mode and one stationary mode. 
The smaller the mass parameter $m_\phi$ with respect to
$\bar{m}_\phi$, the better the straight line approximation. 

\noindent \emph{iv)} If we take $\phi_c \ll
M_P$, we then have the relation, $\bar{m}_{\phi}/H_0 = \bar{m}_N/H_0
\simeq M_P/N_0 \gg 1$, with $H_0$ being the value of the Hubble constant at
the end of inflation. This allows for many oscillations of each field
in a Hubble time, thus the motion is not very suppressed by the
expansion of the Universe and we should expect large amplitudes of
oscillations. Consequently the large oscillations for a reasonable
length of time will allow for the possibility of parametric resonance and
subsequent particle production which must be investigated.

\begin{figure}
\begin{tabular}{cc}
\epsfxsize=8cm
\epsfysize=8cm
\hfil \epsfbox{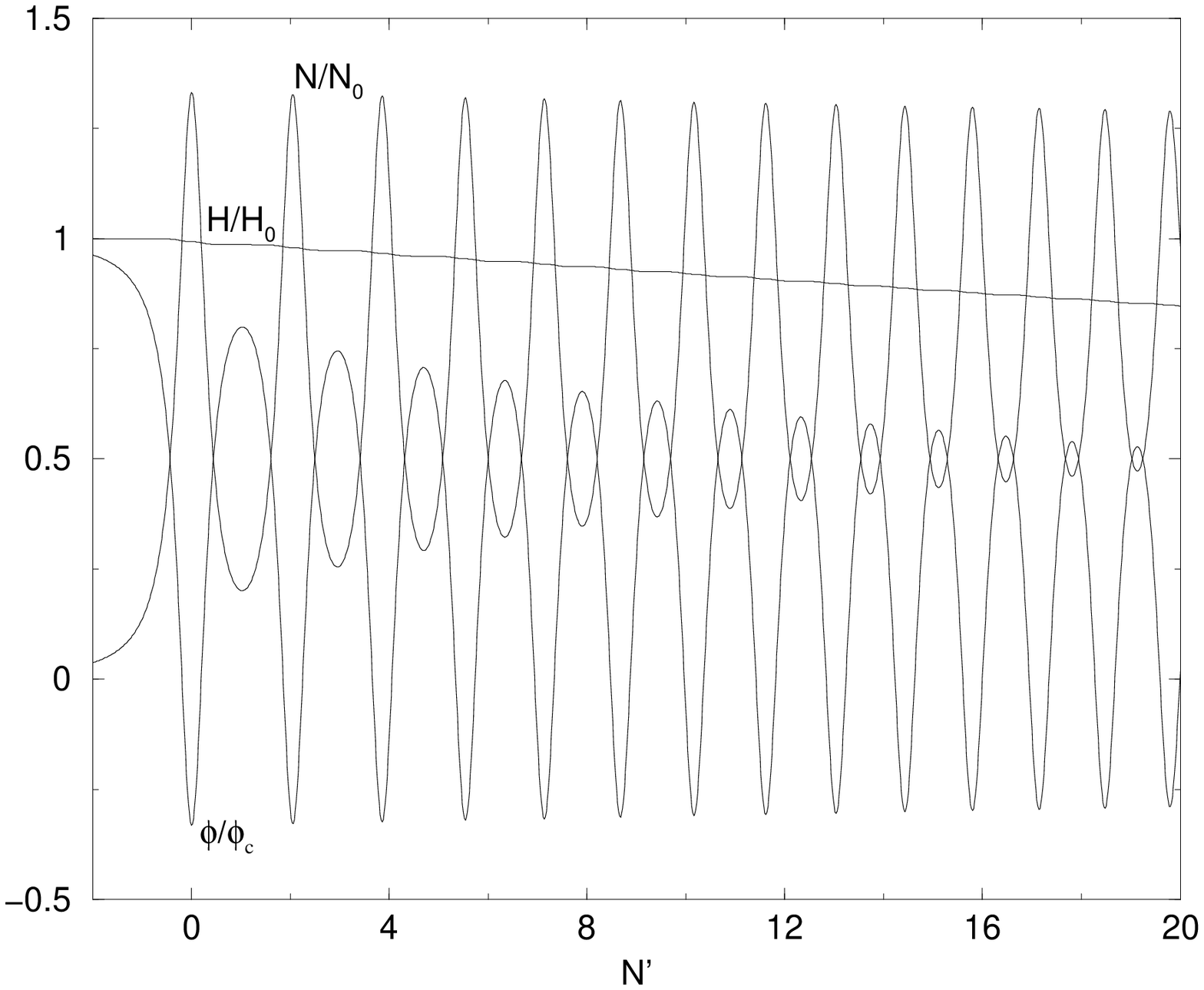} \hfil
&
\epsfxsize=8cm
\epsfysize=8cm
\hfil\epsfbox{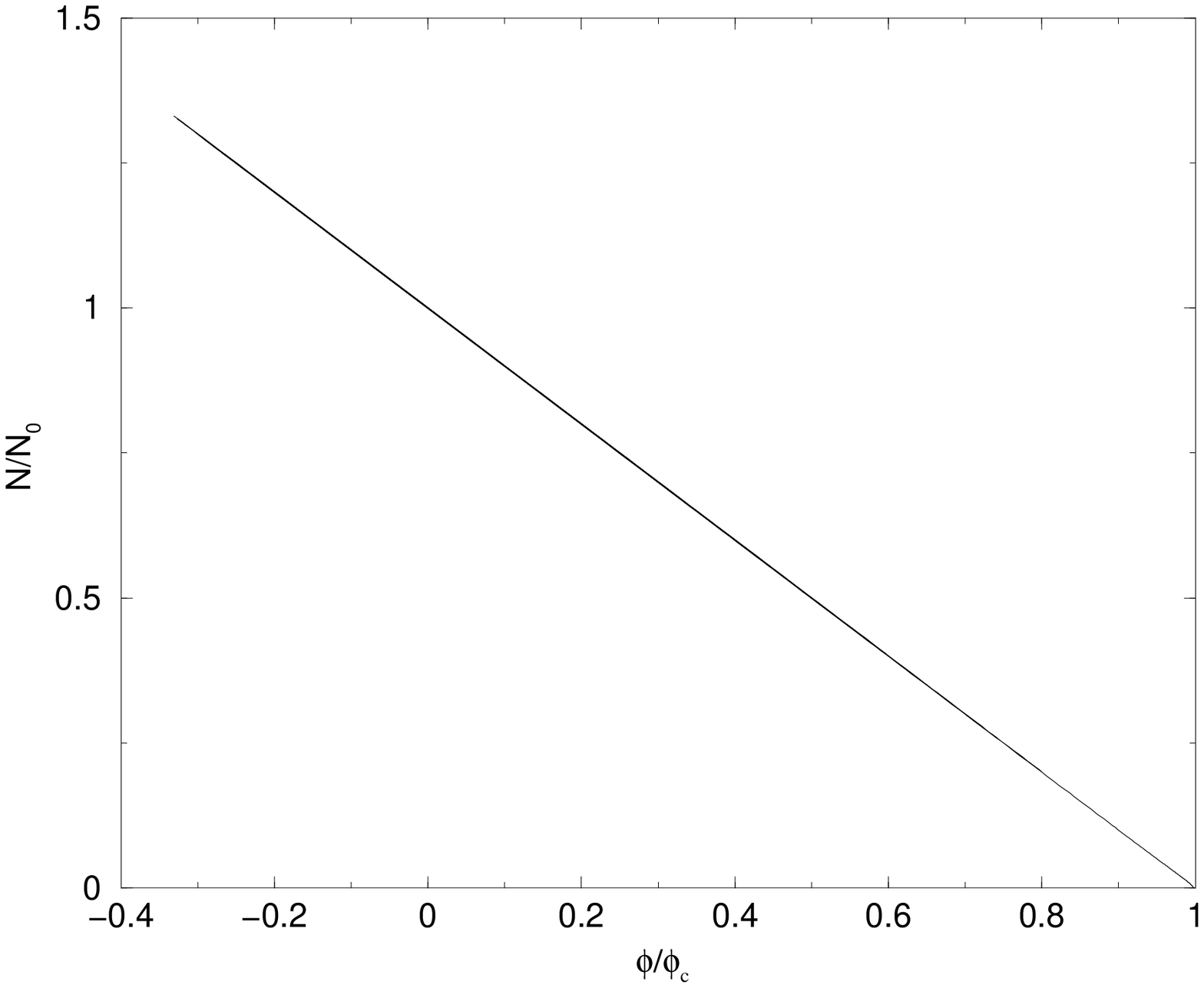} \hfil
\\
\epsfxsize=8cm
\epsfysize=8cm
\hfil \epsfbox{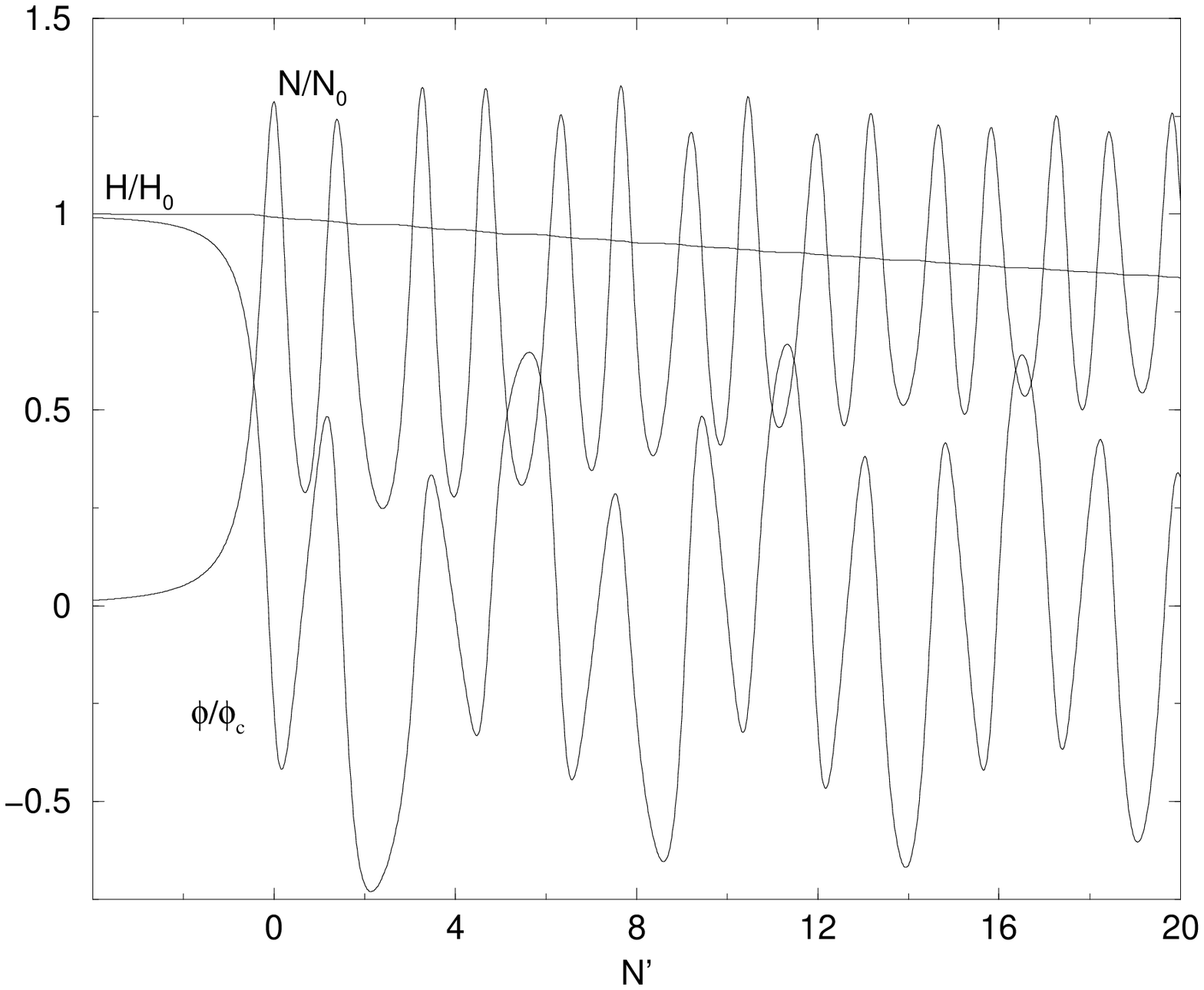} \hfil
&
\epsfxsize=8cm
\epsfysize=8cm
\hfil \epsfbox{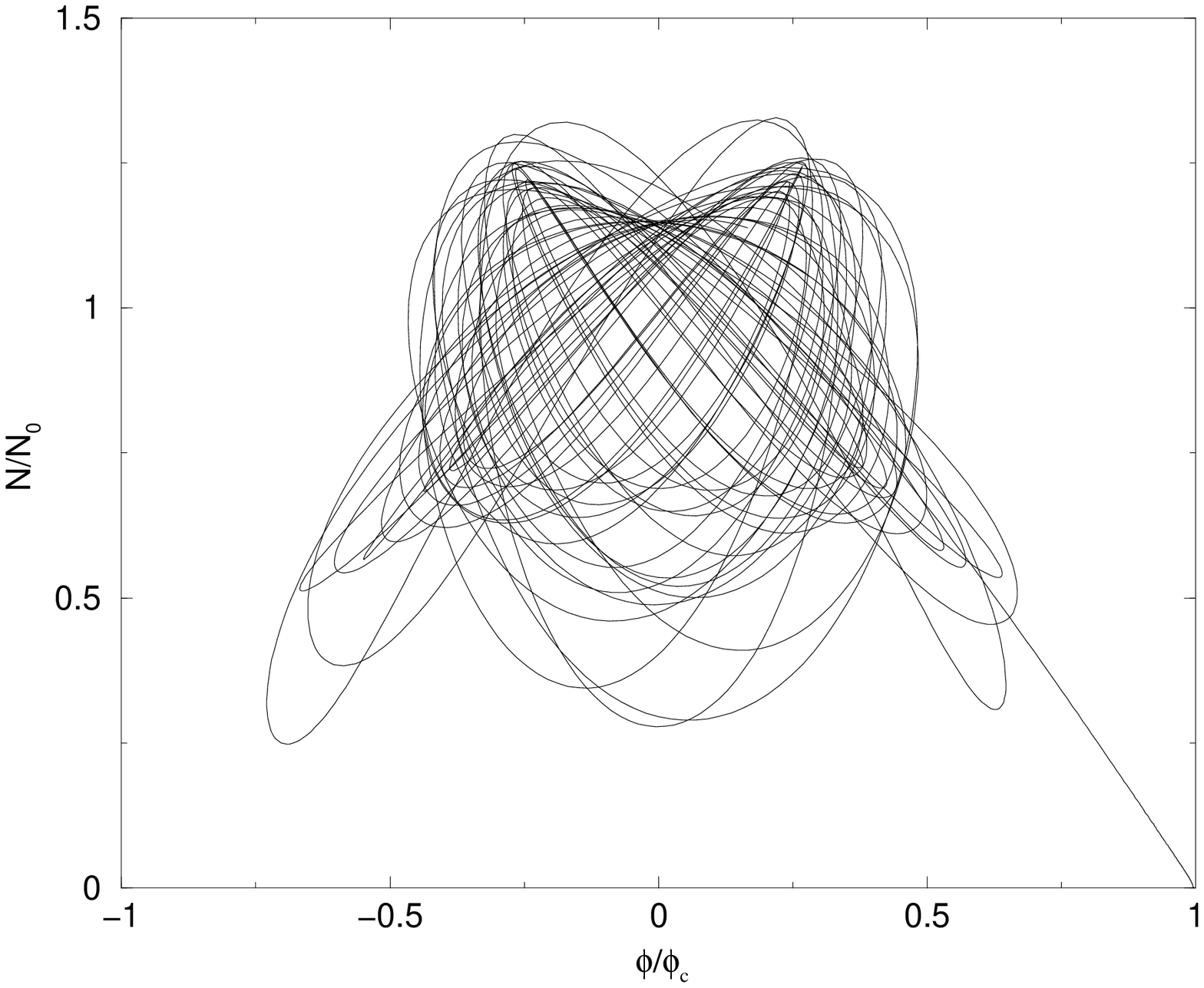}
\hfil 
\end{tabular}
\caption{Oscillations of the classical fields (left) and the
trajectory in the $N-\phi$ plane (right). The top panels correspond to
the supersymmetric hybrid model with $\phi_c=10^{16}\, GeV$,
$\kappa=10^{-3}$ and $m_\phi=3.7\times 10^{9}\, GeV$. The time scale is
given in terms of the approximate number of oscillations of the fields,
$N'=\bar{m}_\phi t /2 \pi$, where $N'=0$ has been defined for
convenience as the point where the fields reach the first peak. The
lower panels show for comparison the situation in a non-supersymmetric
hybrid model with $g= \sqrt{\lambda} = 10^{-3}$, and 
$\phi_c (=\Lambda)$ and $m_\phi$ as before. \label{sugraplot}}
\end{figure}

In Ref. \cite{ref4} they have studied standard hybrid inflation with
$\lambda=g^2$ and found chaotic behaviour of the classical fields
initially, that eventually becomes more regular. However, the SUGRA
hybrid inflation model would correspond to $\lambda= g^2/2$, where the
chaotic motion is not present and the trajectory is given by a
straight line. This can be seen in Fig. (\ref{sugraplot}), where we
have plotted the oscillations of the classical fields and the
trajectory in the $N-\phi$ plane, for the choice of parameters:
$\phi_c= 10^{16}\,GeV$, $\kappa=10^{-3}$ and $m_\phi=3.7\times
10^{9}\, GeV$, which correspond to the choice $\eta \simeq 0.01$. We
have also included for comparison the standard hybrid model with
$\sqrt{\lambda} = g = 10^{-3}$, and the same values of $\phi_c$ and
$m_\phi$. In both situations, the results will remain qualitatively
the same when changing the value of $\phi_c$, and accordingly those of
$\kappa$ and $m_\phi$; the main effect of reducing (increasing) the
scale $\phi_c$ being a weaker (stronger) effect of the expansion of
the Universe during the first oscillations.
 
Using the relation Eq. (\ref{traj}), we can effectively replace the
two oscillating fields problem by a single field, for example $N(t)$,
oscillating along the potential,
\begin{equation}
V(N)= \frac{1}{4} \kappa^2 N_0^4 + \frac{3}{4} \kappa^2 N^4 - \kappa^2
N^3 N_0.
\label{Vsingle}
\end{equation}
Near the origin the potential is almost flat, with the curvature
becoming first negative and then changing sign and increasing as we
approach and pass the minimum. Therefore, the field will tend to spend
most of its time near the initial values, mainly varying when it
approaches the minimum and the evolution speeds up.  In a
non-expanding Universe, and taking as initial conditions the
asymptotic limit $N(-\infty)=\dot{N}(-\infty)=0$, the exact solution
of the classical evolution equation is given by:
\begin{eqnarray}
\frac{N(t)}{N_0} &=& \frac{4/3}{1+2 \bar{m}_\phi^2 t^2/3}
\nonumber \\
& = & 1 + \frac{1}{3}\cos (\int^t_0 
\sqrt{\frac{3}{2}} \left( \frac{N(t')}{N_0} \right) \bar{m}_\phi dt') -
 \frac{2}{3} \sin^2 
(\int^t_0 
\sqrt{\frac{3}{8}} \left( \frac{N(t')}{N_0} \right) \bar{m}_\phi dt') \,,
\label{soln} 
\end{eqnarray}
where $t=0$ is defined when $N$ reaches the peak of the oscillation
$N(0)/N_0=4/3$. In fact the motion can be decomposed into two
different parts, which we label region I/II respectively:

\noindent Region I: when $|N/N_0-1|\le 1/3$ the field is near the
minimum $N_0$, the motion is more harmonic and the sine squared term
in the right hand side of Eq. (\ref{soln}) can be neglected, i.e.,
\begin{equation}
\frac{N(t)}{N_0} \simeq 1 + \frac{1}{3}\cos (\int^t_0 
\sqrt{\frac{3}{2}} \left( \frac{N(t')}{N_0} \right) \bar{m}_\phi dt')\,.
\label{solnmin} 
\end{equation}

\noindent Region II: when $0\le N/N_0 <2/3$ the field
increases/decreases as $N(t) \propto 1/t^2$.

In an expanding Universe, the main effect of the friction term in the
evolution equation will be to reduce the amplitude of the successive
oscillations (and thus the period), such that the amplitude will now
decrease as the inverse of time. Then, we can approximate the behaviour
of the $N$ field around each peak of the oscillations (region I) by,
\begin{equation}
\frac{N(t)}{N_0} \simeq 1 + \frac{\Phi(t)}{3}\cos (\int^t_0 
\sqrt{\frac{3}{2}} \left( \frac{N(t')}{N_0} \right) \bar{m}_\phi dt')\,,
\label{solnminh} 
\end{equation}
where now $\Phi(t) \propto 1/t$. The actual decrease in amplitude over
an oscillation period will depend on the ratio
$H/\bar{m}_\phi$. However, during the first stages of the oscillations
the dominant effect will be to render the total oscillation more
harmonic (symmetric) around the minimum, more than to reduce the
factor $\Phi(t)$.

\subsection{ Particle production: $\delta \phi$, $\delta N$. \label{pp}} 

Due to the behaviour of the classical fields, we may expect
particle production through parametric resonance when the field is
$oscillating$ around the peaks, that is, in region I when $N(t)/N_0 \ge
2/3$. Therefore, in that interval of time we can use the
approximation given in Eq. (\ref{solnminh})  
in order to convert the evolution equations for the quantum
fluctuations, Eq. (\ref{queom}), into a Mathieu-like equation. Here we will
be mainly concerned about the production of $\phi$ and $N$ particles,
which indeed will occur in any SUSY model of hybrid
inflation. Production of other species of particle will be model dependent,
depending on how they couple to the classical fields. 

We first analyse the equations neglecting the mixing terms,
$\mathcal{M}^2_{\alpha \beta}$ in Eq. (\ref{queom}). As before, we will
use the fact that $\phi$ and $N$ follow a straight 
line trajectory, and work only with the classical field $N(t)$. The
effective masses for the quantum fluctuations $\delta \phi$ and
$\delta N$ are then given by:
\begin{eqnarray}
\mathcal{M}^2_{\phi\phi}&=& 2 \kappa^2 N^2  \,,\\
\mathcal{M}^2_{NN} &=& 2 \kappa^2 N ( 2 N - N_0)  \,, \label{mnn}
\end{eqnarray}
which, when the classical fields are near the minimum, become:
\begin{eqnarray}
\mathcal{M}^2_{\phi\phi}& \simeq & \bar{m}_\phi^2 \left( 1 +
\frac{2 \Phi(t)}{3} \cos \int^t_0 \sqrt{\frac{3}{2}} \left(
\frac{N(t')}{N_0} \right)  \bar{m}_\phi dt' \right) \,,\\ 
\mathcal{M}^2_{NN} & \simeq & \bar{m}_\phi^2
\left( 1 + \Phi(t)\cos \int^t_0 \sqrt{\frac{3}{2}} \left( \frac{N(t')}{N_0}
\right) \bar{m}_\phi dt' 
\right) \,.
\end{eqnarray}
We have kept only the contribution linear in the cosine and neglected
the sub-dominant cosine squared contributions. Substituting these
expressions into Eq. (\ref{psieq}), and defining:
\begin{equation}
2 z = \int^t_0 \sqrt{\frac{3}{2}} \left( \frac{N(t')}{N_0}
\right) \bar{m}_\phi dt' \,,
\end{equation}
it is straightforward to get the parameters 
$A_\alpha$ and $q_\alpha$ entering in the Mathieu equation, that is,
\begin{eqnarray} 
q_{\phi}(t)&=&\frac{8}{9} \left(\frac{N_0}{N(t)}\right)^2
\Phi(t),\;\;\;\; 
A_{\phi}(t)= \frac{8}{3} \left(\frac{N_0}{N(t)}\right)^2 \frac{k^2}{(a^2
\bar{m}_\phi^2)}+ 3 \frac{q_\phi(t)}{\Phi(t)},\\
q_{N}(t)&=& \frac{4}{3} \left(\frac{N_0}{N(t)}\right)^2 \Phi(t),\;\;\;\;
A_{N}(t)= \frac{8}{3} \left(\frac{N_0}{N(t)}\right)^2 \frac{k^2}{(a^2
\bar{m}_\phi^2)}+ 2 \frac{q_N(t)}{\Phi(t)}.
\end{eqnarray}
We notice that these parameters are functions of time, as is the case
in models studied in an expanding Universe.  The time dependence
purely due to the expansion appears in the scale factor, $a(t)$, such
that the momentum $\textbf{k}$ is red-shifted as the system evolves, and in the
damping of the amplitude of the classical oscillations, $\Phi(t)$,
which will cause $q(t)$ to decrease appreciably in a Hubble time. On
top of this, we have an extra time dependence related to the nature
of the oscillations, which enters in the factor $N_0/N(t)$, and which
would be present even in Minkowski space.

\begin{figure}
\hfil
\scalebox{0.5}{\rotatebox{270}{\includegraphics{{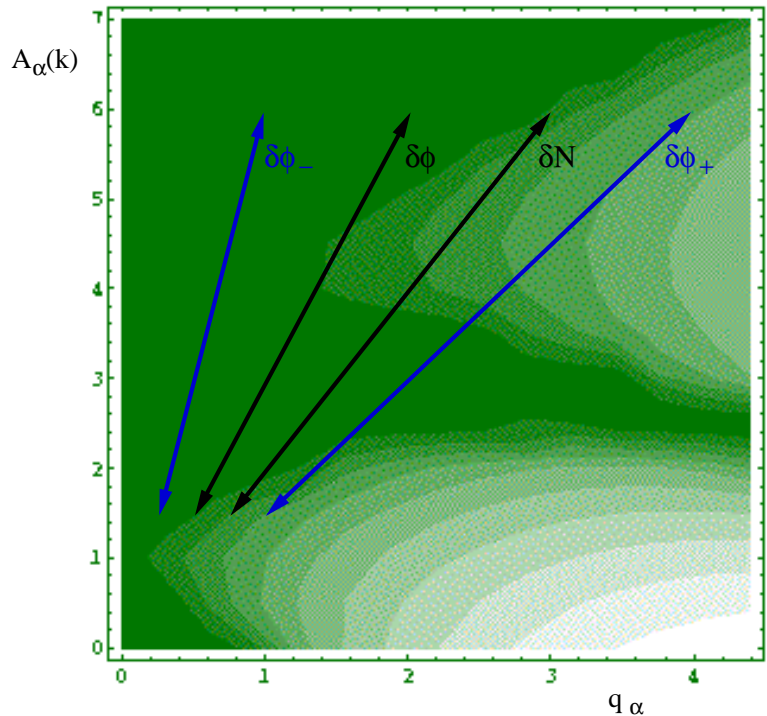}}}} \hfil
\caption{The trajectories through the Mathieu stability/instability
chart for $\textbf{k}=0$ showing how the efficiency of parametric
resonance varies over each stage in the classical field
oscillation. The contours are incremented in levels of $\mu\sim 0.1$,
ranging from the darkly shaded (stable) regions to the light
(extremely unstable) regions. Modes $\textbf{k}> 0$ will lie
vertically above the zero mode trajectories shown. Particle production
occurs mostly in the second instability band. \label{matplot}}
\end{figure}

Although the Mathieu parameters $A_\alpha$ and $q_\alpha$ depend on
time we can infer the behaviour of the solutions using the
stability/instability bands analysis of the canonical Mathieu
equation, (\ref{mat}). For a given mode $\textbf{k}$, instead of
looking to a given point in the chart, we will move along some
trajectory in the plane $(A_\alpha(t), q_\alpha(t))$
\cite{ref7,steinhardt}, crossing different stability/instability
regions as the fields oscillate. As long as we know the trajectory
followed in the Mathieu chart, the standard analysis can be
applied. For example, the parameters $q(t)$ are bounded by,
\begin{eqnarray}
 2  \Phi(t) \ge & q_\phi(t)& \ge  \Phi(t)/2  \,, \\ \nonumber
 3 \Phi(t) \ge  & q_N(t)& \ge \frac{3}{4} \Phi(t) \,,
\label{rangeq}
\end{eqnarray}
so that in each oscillation we will move in the Mathieu plot downward
from right to left during the first half period, and back again. This
can be seen in Fig. (\ref{matplot}).

We first notice that for small values of the momentum $\textbf{k}$ and
$\Phi(t) \approx 1$, we have that $A_\phi \simeq 3 q_\phi$ but $A_N
\simeq 2 q_N$.  The line $A=2 q$ roughly separates the broad (to the
right of this line in the Mathieu plot) from the narrow (to the left)
resonance regimes, and accordingly we expect much less production of
$\phi$ particles than $N$ particles.  To see the difference between
broad/narrow regimes, we can follow Ref. \cite{ref7} in order to get
the maximum $allowed$ growth index in a trajectory of the type $A(k) =
d(k) + b q$, with $k$ being the momentum. This is given by,
\begin{equation}
\mu^{max} = \frac{1}{\pi} \ln \left( \sqrt{ 1 + e^{-\pi c}} + e^{-\pi c}
\right) \label{mumax}\,,
\end{equation}
with:,
\begin{equation}
c= \frac{A(k)-2 q}{2 \sqrt{q}} \,.
\end{equation}
The value $\mu^{max}$ should be understood as  an upper bound on the effective
growth index along the given trajectory. Along the line $A= 2 q$
(i.e. $k=0$) 
we have a constant  maximum possible growth index, $\mu^{max} \approx
0.28$  \cite{ref2,ref7}. Because we are crossing several
stability/instability bands in only half a period, it has been argued
than the effective $\mu$ will somehow be less than half this value,
with the exact value depending on the actual time dependence
of the parameters, and how much time is spent in each band. To
the left of this trajectory, $A > 2 q$, we have that $c$ grows with
$q$ and then $\mu^{max}$ decreases as we move from left to right in the
Mathieu plot. In the opposite case, $A < 2 q$, we have the reverse 
behaviour with  $\mu^{max}$ increasing with large $q$, and we are in
the broad resonance regime.   

\begin{figure}
\epsfxsize=8cm
\epsfysize=8cm
\hfil \epsfbox{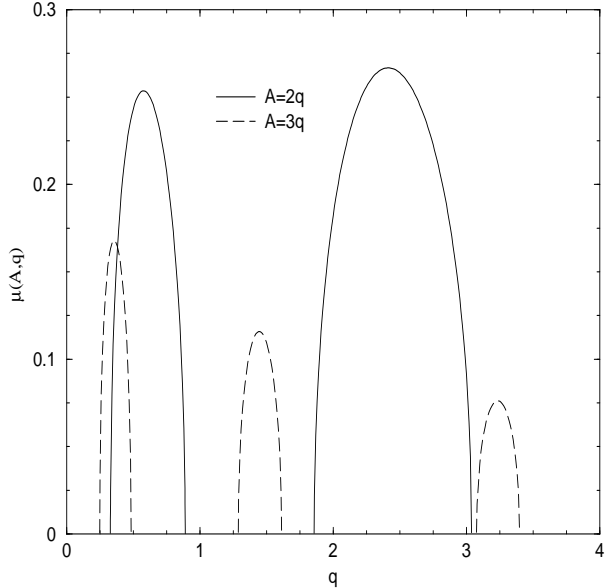} \hfil
\caption{The growth index $\mu(A,q)$ for the static Mathieu
Equation. \label{muindex}}
\end{figure}

Then, it is clear that $N$ production is in the border line between
the broad and narrow resonance, and the resonance can be quite
efficient in producing them. With respect to $\phi$ production, we
only get a non negligible value of $\mu^{max}_\phi \sim 0.1$ in a
small part of the trajectory, as can be seen in Figs. (\ref{matplot})
and (\ref{muindex}). Given the
constraints in Eq. (\ref{rangeq}), at the beginning $\phi$ production
occurs in the second resonance band, whilst $N$ may partially enter
the first resonance band (see Fig. (\ref{muindex})).  Moreover, even
if the motion is not very suppressed by the Hubble expansion,
parametric resonance takes place only during a fraction of the total
period of the motion, so the final effective $\mu_k$ parameter averaging
over several oscillations will be much smaller. Hence, for the $\phi$
particles we expect a narrow resonance if at all for momenta $k \lsim
\bar{m}_\phi$. As the damping factor $\Phi(t)$ gets reduced in time,
$N$ production will be shifted further to the left in the Mathieu chart, into
the narrow resonance regime.

However, $N$ particles with low momenta may be produced when the
classical fields are not near the minimum, although this has nothing
to do with parametric resonance. Whenever $N(t)/N_0 < 1/2$ the
effective squared mass of the $N$ field, Eq. (\ref{mnn}), becomes
negative and we enter the spinodal instability region. In this region,
for the modes for which $k^2/a^2 + \mathcal{M}^2_{NN} < 0$, we will
again have an exponential increasing solution that can be interpreted
as production of particles. At the end of the day, it seems that
production of $N$ particles is quite efficient for modes with $k \lsim
\bar{m}_\phi$, and extremely effective for small modes below roughly
$0.35 \bar{m}_\phi$.

So far, so good. These would be the qualitative results if the
$mixing$,  $\mathcal{M}^2_{\phi N}$, were negligible, but this is
never the case in SUSY hybrid inflation. The analysis with
the mixing term for the $\delta \phi$, $\delta N$ system is greatly
simplified given the fact that the background fields follow a
straight line trajectory, implying only one oscillating
mode. Hence, the coupled system for their quantum 
fluctuations can also be decomposed into two orthogonal modes,
\begin{eqnarray*}
\delta \phi_+ &=& \cos \theta \delta \phi + \sin \theta \delta N \,, \\
\delta \phi_- &=& -\sin \theta \delta \phi + \cos \theta \delta N \,,
\end{eqnarray*}
with $\tan \theta = - \sqrt{2}$. The time dependent frequencies for
the modes $\delta \phi_{\pm}$ are now given by:
\begin{eqnarray}
\mathcal{M}_{+}^2&=& 2 \kappa^2 N(t) ( 3 N(t) - 2 N_0) \,, \\
\mathcal{M}_{-}^2&=& 2 \kappa^2 N(t) N_0 \,,
\end{eqnarray}
and  the corresponding parameters in the Mathieu equation are:
\begin{eqnarray}
q_{+}(t) &=& \frac{16}{9} \left(\frac{N_0}{N(t)}\right)^2 \Phi(t),\;\;\;\;
A_{+}(t)=\frac{8}{3} \left(\frac{N_0}{N(t)}\right)^2 \frac{k^2}{(a^2
\bar{m}_\phi^2)}+ \frac{3}{2} \frac{q_{+}(t)}{\Phi(t)} \,, \\
q_-(t) &=&\frac{4}{9} \left(\frac{N_0}{N(t)}\right)^2 \Phi(t),\;\;\;\;
A_-(t)= \frac{8}{3} \left(\frac{N_0}{N(t)}\right)^2 \frac{k^2}{(a^2
\bar{m}_\phi^2)}+ 6 \frac{q_{-}(t)}{\Phi(t)}\,. 
\end{eqnarray}
The Mathieu parameters for the mode $\delta \phi_{-}$ appear well to
the left of the line $A=2 q$, i.e. deep in the narrow resonance
regime (see Fig. (\ref{matplot}). However, for the mode $\delta
\phi_+$ we have the reverse 
situation, and the amplification of this mode will occur in the
efficient broad resonance regime. Moreover, the effective mass
$\mathcal{M}^2_{+}$ becomes negative when $N(t)/N_0 < 2/3$, and again
we will also have production of particles in region II of the
classical oscillation.

Finally, given that $\tan \theta \approx
O(1) $, both $\delta \phi$ and $\delta N$ are dominated by the
production in the $\delta \phi_+$ mode, 
\begin{equation}
\delta \phi \simeq - \sin \theta \delta \phi_+ \,,\;\;\;\;\; 
\delta N \simeq \cos \theta \delta \phi_+ \,,
\end{equation}
leading to roughly the same rate of production for both
particles. Moreover, in this case the mixing enhances the rate of
production of $N$ particles compared with what was expected without
it. This enhancement is a two-fold effect: on one hand parametric
resonance is more effective due to the mixing, on the other hand we
have a greater negative frequency in region II, with
$\mathcal{M}^2_+|_{min}= -\bar{m}^2_\phi /3$, which will imply a wider
range of momenta, $k/a < 0.58 \bar{m}_\phi$, in which particles can be
more efficiently produced.    

\begin{figure}
\begin{tabular}{cc}
\epsfxsize=8cm
\epsfysize=8cm
\hfil \epsfbox{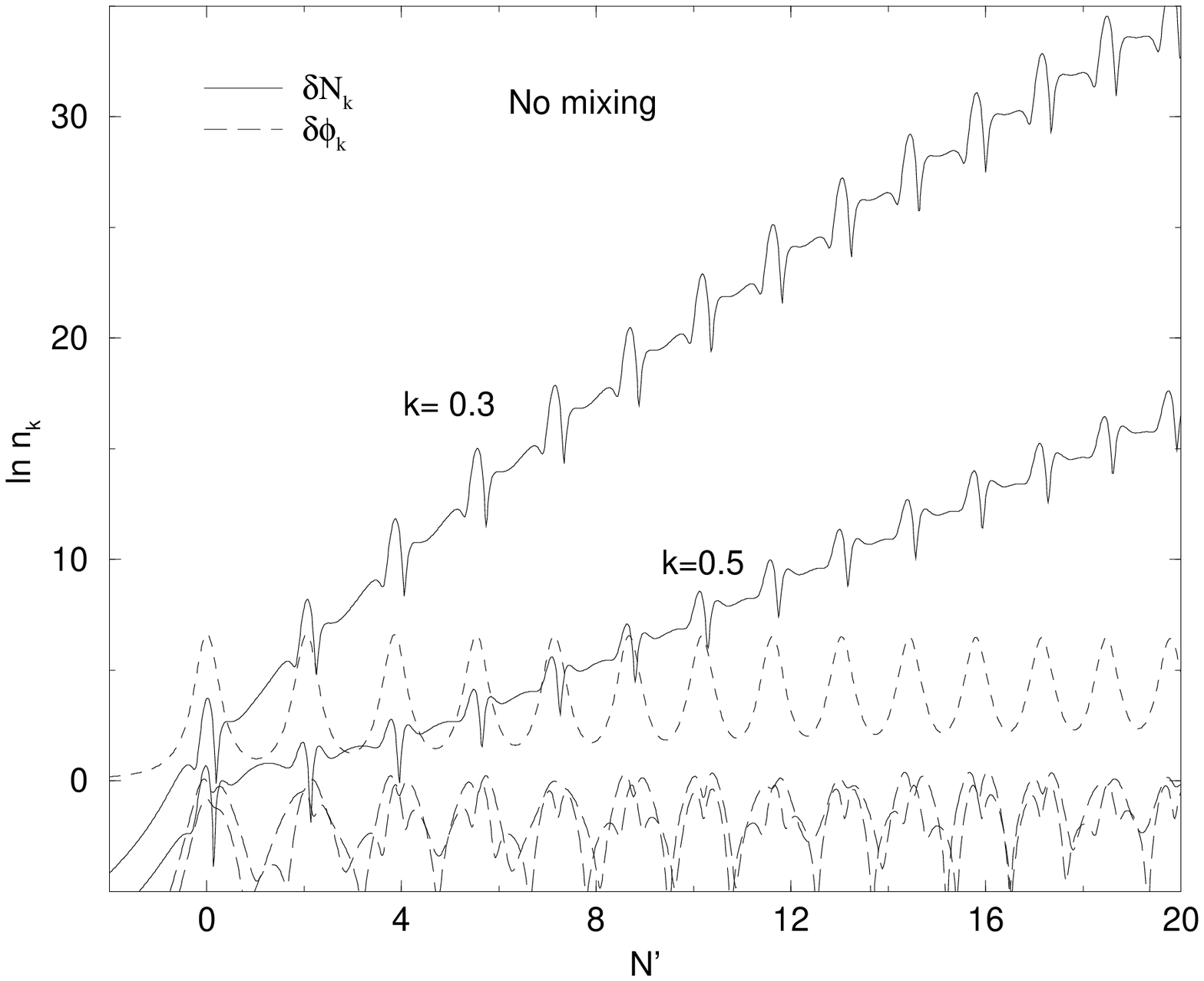} \hfil
&
\epsfxsize=8cm
\epsfysize=8cm
\hfil \epsfbox{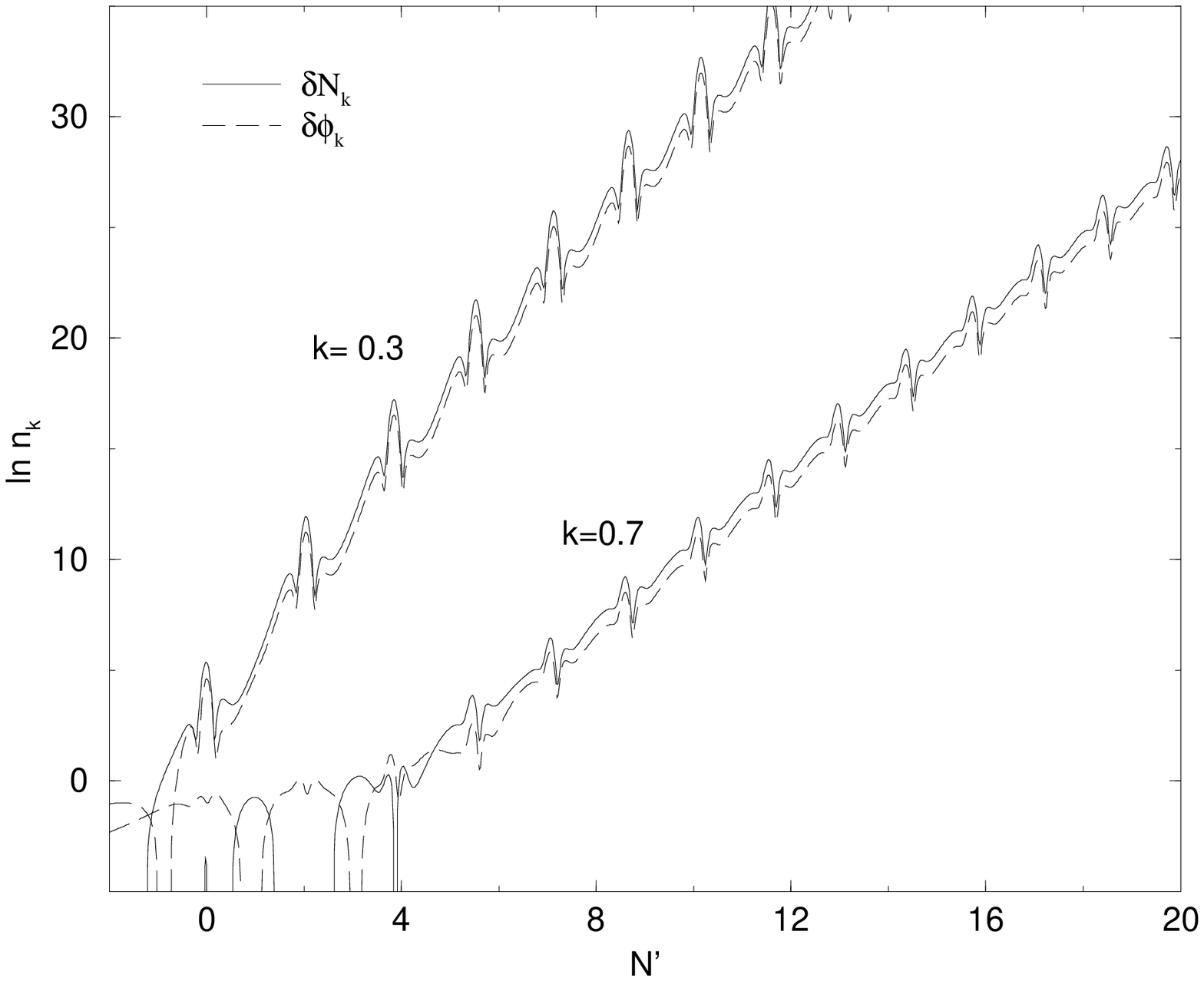} \hfil
\end{tabular}
\caption{Initial rates of production of $\phi$ and $N$ particles for
the model with $\kappa=10^{-3}$, $\phi_c= 10^{16}\, GeV$ and $m_\phi=
3.7 \times 10^{9} \, GeV$. In the left side plot the results without
taking into account the mixing term in the effective squared mass
matrix are displayed; we have also included the oscillation of the
classical field $N(t)$ (dashed line) scaled by a factor of 5. The
comoving momentum $k$ is given in both plots in units of the natural
frequency $\bar{m}_\phi$ and it corresponds to the physical momentum
at $N' \approx -17$. In the right side plot, by the time particles are
produced the initial value of the physical momentum $k/a=0.7
\bar{m}_\phi$ has been red-shifted below $0.58 \bar{m}_\phi$.
\label{shybln}}
\end{figure}

To illustrate these points, we have numerically integrated the
equations with $\kappa=10^{-3}$, $\phi_c= 10^{16}\, GeV$ and
$m_\phi=3.7 \times 10^{9}\, GeV$, and obtained the occupation number
of particles, Eq. (\ref{occno}). The results are plotted in
Fig. (\ref{shybln}), and we have included for comparison those
obtained when neglecting the mixing (left plot). Without the mixing,
$\phi$ particles are hardly produced, whilst production of $N$
particles with low momenta proceeds quite quickly. Moreover, we can
distinguish clearly in the plots the two different mechanisms of
particle production. The peaks in $\ln n_k$ are due to parametric
resonance in an expanding Universe, whereas between peaks $\ln n_k$
grows almost linearly in the first few oscillations reflecting the
presence of a negative squared frequency in the evolution equation for
these modes. When the Hubble expansion has reduced the amplitude of
the oscillation such that $N(t)/N_0 > 1/2$ (without mixing) or further
to $N(t)/N_0 > 2/3$ (with mixing), the occupation numbers become
constant between the peaks. When the mixing is included (right plot),
the occupation numbers for both particles become comparable, and the
rate of production of $N$ particles is enhanced compared to the
situation without the mixing.

\subsection{ Comment on backreaction. \label{comment}}

Up to now, the analysis of particle production in the SUSY hybrid
model has been done neglecting backreaction effects. Even if no other
particles couple to the inflaton and $N$ fields, it can be inferred
from Fig. (\ref{shybln}) that the production of their own quanta will
soon affect the evolution of the system. Due to the presence of the
tachyonic mass immediately after the end of inflation, these fluctuations
may grow and modify the classical evolution even before the fields
reach the minimum for the first time. After the inflaton has crossed
the critical value, the classical fields may still move in the
slow-rolling regime for a while, and we can consider that the Universe
continues inflating for a further fraction of an e-fold $N_e \approx H
t$ (for the parameters of Fig. (\ref{sugraplot}) $N_e \approx
0.3$). The first effect of the instabilities during this period would
be to speed up the evolution and end the slow roll regime and thus
inflation altogether. Their contributions $\langle \delta
\phi_\alpha^2 \rangle$ to the classical equations can be computed
using the Hartree Fock approximation\footnote{A brief description of
this approximation is given later in section \ref{hartree} and
Appendix B.} and can be interpreted as corrections to the effective
masses of the fields. Thus, even in the slow roll regime with $N(t)
\approx 0$, the mass of the inflaton becomes,
\begin{equation}
m_\phi^2  \rightarrow m_\phi^2 + 2 \kappa^2 \langle \delta N^2 \rangle
\,.
\end{equation}
When the fluctuations grow such that both terms become comparable, the
inflaton (and by extension the $N$ field) is pushed away from the slow
roll trajectory. These contributions are dominated by the modes with
the lower values of the momentum $\textbf{k}$, but their evolution
will tend to follow that of the background fields (for small enough
values of $\textbf{k}$, the equations of motion are the same). As
the classical fields reach the minimum and begin to oscillate,
production of particles in these low momentum modes practically stops,
and production begins in slightly higher modes.  This effect is similar to the
``zero mode assembly'' described in Ref. \cite{refDan1} in the context
of new inflation type models.  Despite the differences, new inflation
and hybrid inflation share the fact that both end with a phase
transition. Under certain conditions, it is shown how the spinodal
instabilities drive the growth of the quantum fluctuations such that
they dominate completely the inflationary dynamics ending
inflation. The super-horizon modes together with the inflaton follow
the same evolution and assemble into an effective zero-mode field
which drives the dynamics of the system.  In hybrid inflation, the
contribution due to the particles produced in the lower modes during
the pre-oscillatory, quasi slow-rolling regime, speeds up the initial
evolution towards the minimum, but it does not reach the level of
modifying the oscillatory regime. For this to happen, we have to
wait until higher modes have been populated and become dominant.

On the other hand, other particles can be present and these might be
more efficiently produced during the oscillatory phase, such that they
would dominate the backreaction effects. Given that in the original
SUSY superpotential we have complex fields, their imaginary
components, CP-odd singlet scalars, are among the particles that
should also be included in the game. However, because the couplings to
their real partners is model dependent, we will not attempt to study
the production of these particles in a general way here. We will
rather focus in the next section in a particular supersymmetric model
of inflation motivated by particle physics considerations, where these
questions can be addressed.

\section{A particular model: $\phi$NMSSM Hybrid Inflation}

The $\phi$NMSSM \cite{phiNMSSM} is a simple modified version of a well
known particle physics theory, the next-to-minimal supersymmetric
model (NMSSM). As such it is a realistic working supersymmetric model
which supports inflation. The model has been formulated in the context
of a SUGRA theory in \cite{SUGRA}. We will briefly review the global
SUSY model here.

The model is based on the superpotential:
\begin{equation}
W = \lambda \rm{N}\rm{H}_1 \rm{H}_2 - \it{\kappa}\rm{N}^2\Phi + ...,\label{SP}
\end{equation}
where $\rm{H}_1, H_2$ are the standard Higgs doublet superfields, and
N and $\Phi$ are singlet superfields. As in the usual NMSSM, ${\kappa}$ and
$\lambda$ are two dimensionless coupling constants. Since the singlet,
$\Phi$ is linear in the superpotential, the potential will be very
flat in its direction enabling it to act as the slowly rolling
inflation field.

This superpotential admits a global $U(1)_{PQ}$ Peccei-Quinn symmetry
which forbids further terms and solves the strong-CP problem. The
symmetry is broken at the scale of the singlet VEVs giving rise to an
invisible DFSZ axion of that scale. Whilst the axions may provide a
Dark Matter candidate, they must be tightly constrained in order not
to over close the Universe or destroy the results of
nucleosynthesis. Axions may be produced directly during the explosive
preheating phase or they may be a decay product of the other massive
particles produced during preheating.

Previous to the start of inflation it is assumed that the field 
$N$ is driven to zero. During inflation the standard Higgs fields may
be ignored and the scalar potential in terms of the canonically
normalised real fields $\phi$ and $N$, becomes:
\begin{equation}
V_0(\phi,N) = V(0) + \frac{1}{4}{\kappa}^2N^4 + {\kappa}^2(\phi -
\phi_{c-})(\phi - 
\phi_{c+})N^2 + \frac{1}{2}m_\phi^2 \phi^2,  \label{Repotl}
\end{equation}
The constant, $V(0)$ has been added to ensure the necessary vanishing
of the potential at its global minimum which corresponds to our vacuum
and the critical values of $\phi$ for which the effective $N$ mass
vanishes have been written explicitly in the potential and are given
below in terms of the soft SUSY parameters, which appear in the soft
SUSY-breaking part of the potential, $V_{soft}= \frac{1}{2}m_N^2 N^2 -
\frac{1}{\sqrt{2}}\kappa A_\kappa \phi N^2$,
\begin{equation}
\phi_{c\pm} = \frac{A_{\kappa}}{2\sqrt{2}{\kappa}}\left(1 \pm
\sqrt{1-\frac{4m_N^2}{A_{\kappa}^2}}\right). 
\end{equation}
The first derivatives of the potential are,
\begin{eqnarray}
\partial V_0/\partial\phi &=&2{\kappa}^2(\phi-\phi_0)N^2 + m_\phi^2\phi,
\nonumber \\ 
\partial V_0/\partial N &=&{\kappa}^2N[N^2 + 2(\phi-\phi_{c-})(\phi -
\phi_{c+})]. 					            \label{derivs}
\end{eqnarray}
Taking the fields to be positive, the (tree level) global minimum is
found to be,
\begin{eqnarray}
\phi_0 & = & \frac{\phi_{c+}+\phi_{c-}}{2} =
\frac{A_{\kappa}}{2\sqrt{2}{\kappa}}, \nonumber \\ 
N_0 & = & \frac{\phi_{c+}-\phi_{c-}}{\sqrt{2}}
=\frac{A_{\kappa}}{2{\kappa}}\sqrt{1-\frac{4m_N^2}{A_{\kappa}^2}},  
\end{eqnarray}
and thus the vacuum energy which must be zero at the global minimum is
given by,
\begin{equation}
V(0) \approx -V_0(\phi_0, N_0) = \frac{1}{4}{\kappa}^2 N_0^4.
\end{equation}
In the SUGRA model \cite{SUGRA} $V(0)$ is provided in the hidden
sector by the moduli and dilaton contributions.
 
As usual, during inflation the
$N$ field is trapped in its false vacuum, $N=0$, whilst the $\phi$
field slowly rolls along the potential either from above $\phi_{c+}$
towards the origin (hybrid)  or from below $\phi_{c-}$ and away from
the origin (inverted hybrid  \cite{IH}), depending on whether the small
inflaton mass squared, $m_{\phi}^2$, is positive or
negative respectively. 
Once at the critical value, the $N$ mass 
changes sign and both singlets roll down towards the global minimum
signaling the end of inflation. 

The distinctive feature of having two critical values, $\phi_{c\pm}$, 
relies on the presence of the trilinear parameter $A_k$ in the
potential. Since we expect $\lambda N_0 \sim 10^3\, GeV$ in order to
motivate the $\mu$ parameter of the MSSM, the inclusion in the
potential of the SUSY-breaking soft parameters, masses and trilinears,  
is mandatory in our model. However, once the potential is written as
in Eq. (\ref{Repotl}), it can be seen as a generalisation of the 
SUSY hybrid potential, independently of the origin of the mass
parameters. In the limit $A_k^2 \ll |m^2_N| $,($m_N^2 
<0$), and thus $\phi_{c-} = -\phi_{c+}$, we recover indeed the
standard SUSY hybrid potential,    
\begin{equation}
V_0(\phi,N) = \frac{1}{4}{\kappa}^2 (N^2 -2 \phi^2_c)^2 + {\kappa}^2
N^2 \phi^2 + \frac{1}{2} m_\phi^2 \phi^2.  \label{Repoth}
\end{equation}
Therefore, the main difference between Eqs. (\ref{Repotl}) and
(\ref{Repoth}) lies in the origin of the scale  $\phi_c$. 
In standard SUSY F-term hybrid inflation, this is given in terms of
an intermediate mass scale $\Lambda$ coming directly from the
superpotential, Eq. (\ref{supot}), 
whereas in our model no explicit mass scale is included in the
superpotential and the critical value, $\phi_c \approx
A_k/\kappa$, is generated via soft SUSY-breaking. From the cosmological
point of view, this in turn will mark the difference between the
models  through the value of the Hubble constant and the vacuum energy
during inflation. 

Whatever the origin of the critical scale $\phi_c$, the parameters
in the model 
must satisfy the COBE constraints  and the slow rolling conditions,
Eqs. (\ref{epsilon}-\ref{cobecon}). 
Since in our model $\phi_c \sim \phi_0 \sim N_0$ must not exceed the bound on
the axion scale, and moreover $\lambda N_0 \sim \kappa
\phi_c \sim 1 \, TeV$,  
rough order of magnitude estimates
for the parameters in the potential are found to be: $\phi_0 \sim N_0
\sim \phi_c\sim 10^{13}GeV,\; \lambda\sim {\kappa}\sim 10^{-10}$,
hence the vacuum energy and Hubble constant are small, $V(0)^{1/4}\sim
10^8GeV$, $\; H_0 \sim 10^{-3} GeV$. Note that terms involving the
Higgs doublets in the potential, (see Appendix A, Eq. (\ref{fullpotl})),
will be at least a factor $\mathcal{O}({\kappa}^2)$ below the value of
$V(0)$ above so essentially these fields will not contribute at all
towards the vacuum energy. We result with a scale independent
spectrum, $n=1$ predicted to an accuracy of $10^{-12}$. The inflaton
mass, $m_{\phi}\sim eV$ comes from radiative corrections to its zero
tree-level mass. The smallness of $\lambda$ and ${\kappa}$, the origin
of $V(0)$ and the massless tree-level inflaton can be explained in
terms of a no-scale SUGRA, superstring inspired theory in \cite{SUGRA}
where it is also shown that the $\eta$-problem common to most
($F$-term) inflation models can be avoided here.


In the next  subsections we give the results of our investigation of
the post-inflationary stage of the $\phi$NMSSM model described in the
previous section. 
Firstly we analyse the behaviour of the classical homogeneous
background fields according to the classical potential, neglecting
particle production (Figs. \ref{fig:traj}-\ref{fig:Hub}); secondly
we study the quantum fluctuations on the classical background of all
fields in the theory, relating the growth of these fluctuations to
particle production (Figs. \ref{fig:pprod}-\ref{fig:rates}); and
lastly we find an estimate of the point at which backreaction of
created particles becomes important and show the immediate effect this
has on the classical field evolution (Figs. \ref{phiNbr}-\ref{ntbr}).

Within the approximate order of magnitude results necessary to realise the
model, we have chosen the set of parameters below for our numerical
study:
\begin{eqnarray}
\phi_0 = N_0 = 10^{13}GeV,&&{\kappa} = 10^{-10}, m_\phi = 10^{-9}GeV,
\label{params}\\ 
\Rightarrow&&\frac{\phi_{c\pm}}{\phi_0} = 1 \pm
\frac{1}{\sqrt{2}}. \nonumber                              
\end{eqnarray}

More explicit details and calculations involving the other fields in
the potential needed in the analysis of the post-inflation era
are given in the Appendix at the end of the paper.

\subsection{Classical results - neglecting particle
production}\label{sec4.1} 

\begin{figure}
\hfil
\scalebox{0.5}{\rotatebox{270}{\includegraphics{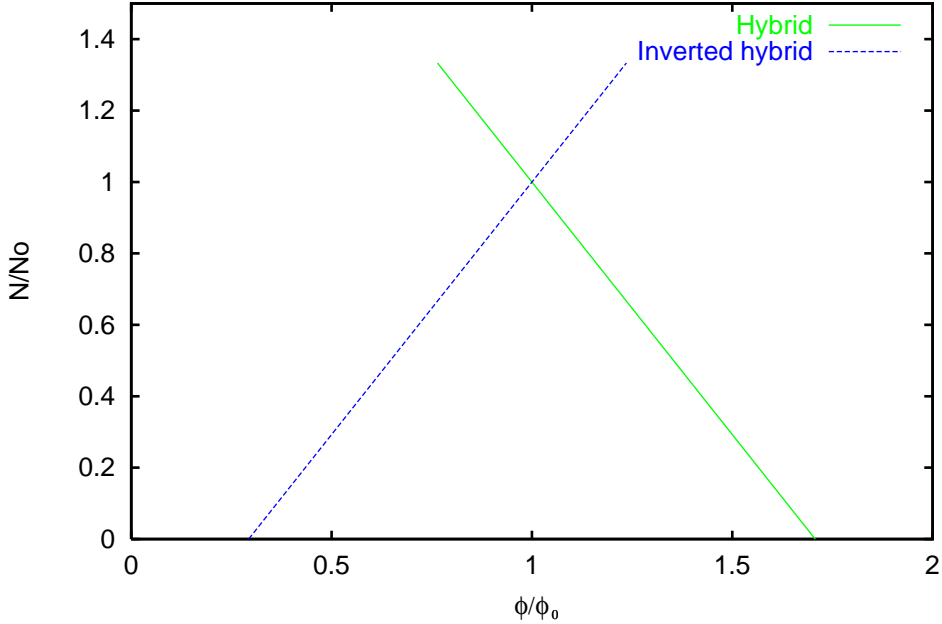}}} \hfil
\caption{Classical trajectory in field space ($\phi,N$) after
inflation, (for the parameters given in (\ref{params})) in the case of
both hybrid and inverted hybrid inflation. The fields begin at
$\phi=(1 \pm 1/\sqrt{2})\phi_0, N=0$, and oscillate around
$\phi=\phi_0, N=N_0$, where, neglecting particle production, they
eventually settle.}
                                                          \label{fig:traj}
\end{figure}

Being no more than a particular version of the SUSY F-term hybrid
model, the $\phi$NMSSM model leads to an evolution of the classical
homogeneous fields, $N(t)$ and $\phi(t)$ (ignoring particle
production) which follows the general pattern given in section
\ref{clshyb}. In this case the inflaton field $\phi$ will oscillate
around its non-zero VEV, $\phi_0$. The two ratios of scales that fix
the behaviour of the oscillations are:

\noindent (a) $m_\phi/ (\kappa \phi_c) \approx 10^{-12}$; the motion
begins with extremely small slow-rolling values, and moreover the
contribution of the tiny inflaton mass $m_\phi$ becomes negligible
very soon in the evolution. As a result, the fields will follow the
straight line trajectory (Fig. (\ref{fig:traj}) to a very good degree of
accuracy, and the actual oscillations of the fields are indeed in
phase.

\noindent (b) $H_0/ (\kappa \phi_c) \approx 10^{-6}$; the oscillations
will begin with very large, lightly damped amplitudes, and oscillate
$\sim 10^{5}$ times in a Hubble time. 

\begin{figure}
\scalebox{0.315}{\rotatebox{270}{\includegraphics{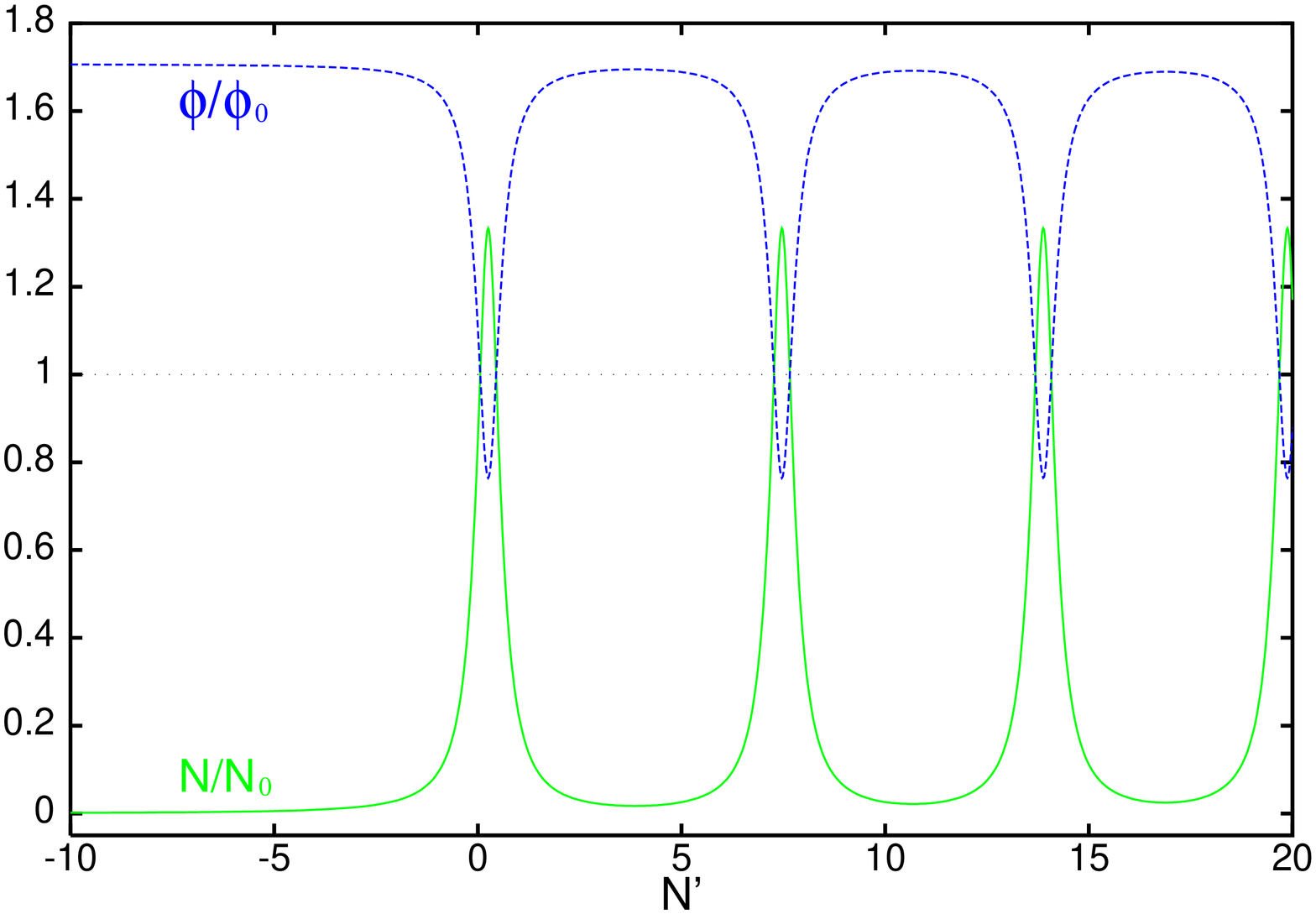}}}
\scalebox{0.315}{\rotatebox{270}{\includegraphics{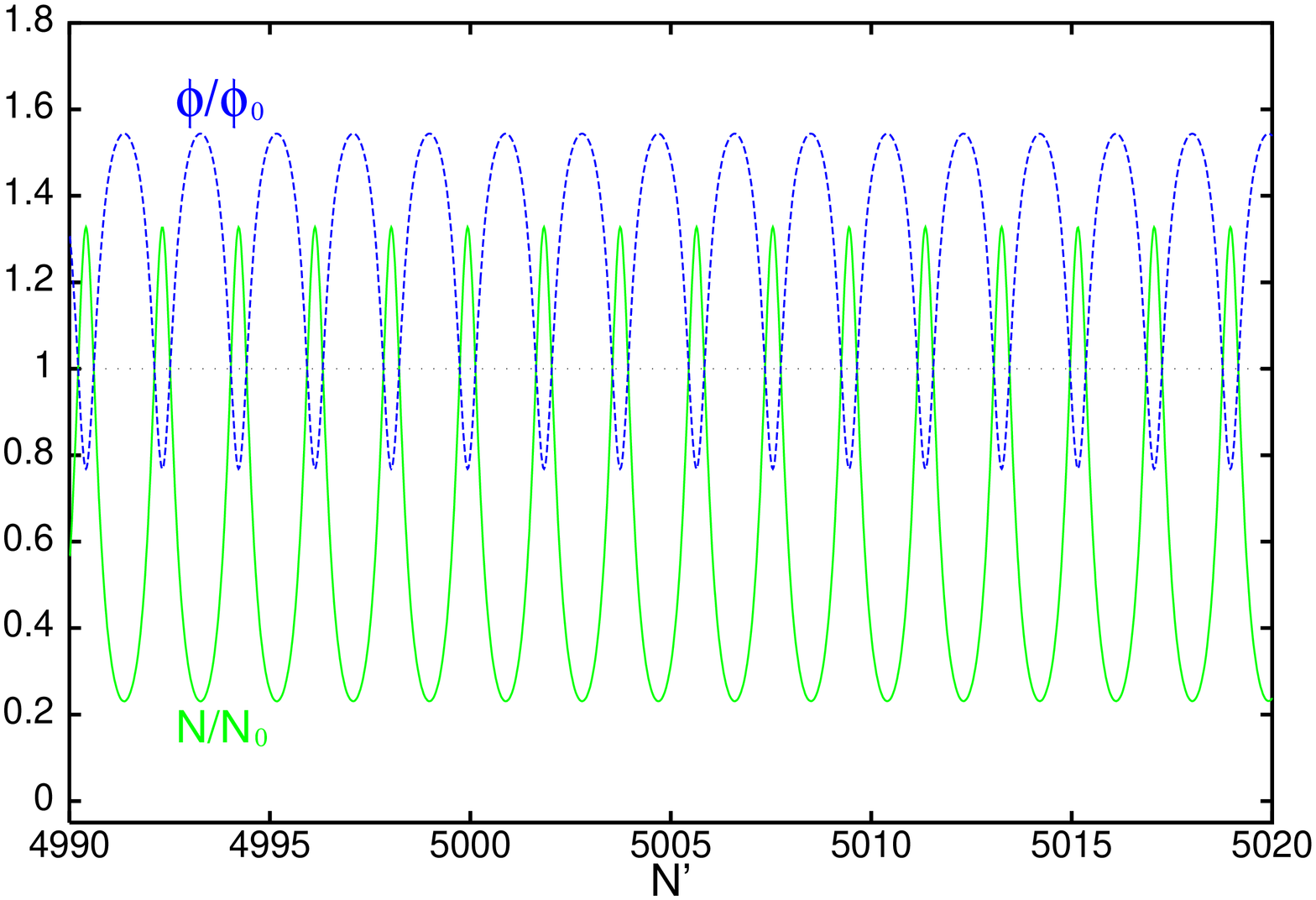}}}
\scalebox{0.315}{\rotatebox{270}{\includegraphics{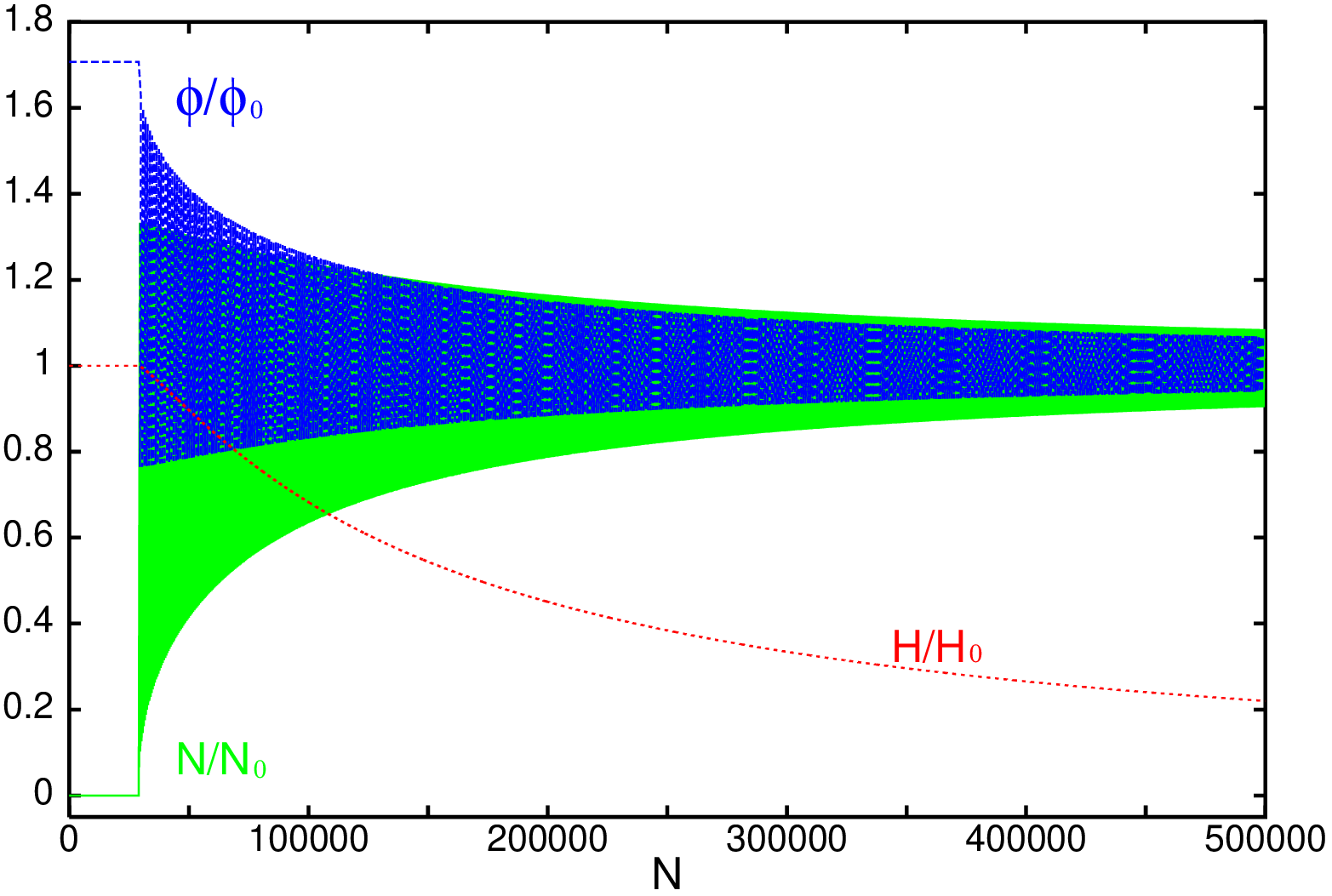}}}
\scalebox{0.315}{\rotatebox{270}{\includegraphics{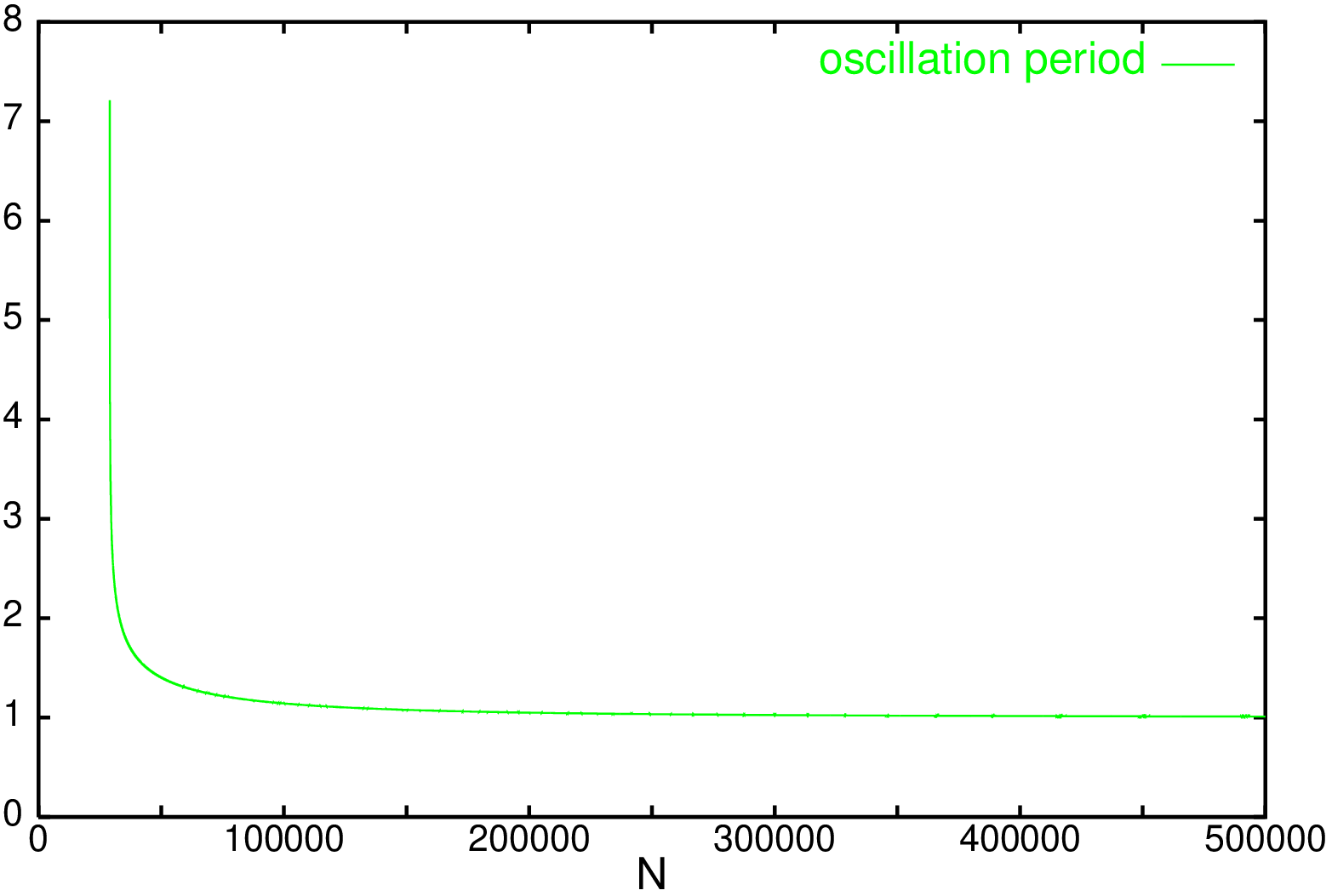}}}
\caption{Top: Oscillations of the classical fields in the hybrid case
at the first stage and at a later stage in their evolution. Bottom
left: Evolution of the fields and Hubble constant after inflation in
the hybrid case over $\sim 2$ e-folds, showing the damping of the
amplitudes of oscillations caused by the decreasing Hubble
constant. The fields take $\sim 0.1$ e-folds before they begin to
oscillate in the shaded regions of the plot. Bottom right: The
decreasing period of oscillations in units of $N$ (period of simple
harmonic oscillations).}
\label{fig:evol}
\end{figure}

Fig. (\ref{fig:evol}) shows the evolution of the fields from the
moment the critical point is passed. The 
time scale displayed is in terms of the approximate number of
oscillations of the fields, given by:
\begin{equation}
N=\frac{\bar{m}_\phi t}{2\pi}=\frac{\bar{m}_N t}{2\pi},
\end{equation}
where one e-fold ($N_e \simeq H t $) corresponds to roughly $\sim 2.7
\times 10^5$ oscillation times. This approximation becomes more exact as the
oscillations approach simple harmonic motion at late times in
the evolution when the amplitudes are very small and the masses given
by (\ref{masses}) become a better approximation to the actual
frequency of these late oscillations. In several of the plots
we have used the variable $N'=N-N_{start}$, where $N'=0$ is defined to
be the moment when the fields reach the global minimum for the first time.

We can see in Fig.~\ref{fig:evol} that the fields take a reasonable
length of time, $N_{start}\sim$~$2.9 \times 10^4$ oscillation times
(about 1/10 of a Hubble time) before they begin to oscillate, so in
fact inflation continues for just a fraction of an e-fold after the
critical point is passed. The fields then begin to oscillate with
amplitudes decreasing due to the expansion of the Universe. In terms
of a Hubble time, the total amplitude will decrease rapidly at
first and more slowly as time progresses until after about 2 e-folds
the oscillations have decreased to $\sim 1/10$th their original
amplitude. The frequency of oscillations increases as they progress,
and the motion eventually evolves towards simple harmonic. 
The period of oscillation rapidly decreases to begin with, indicating that in
comparison to the Hubble time, simple harmonic motion becomes a good
approximation relatively early on in the stages of oscillation.
However, particle production will occur during the first few
oscillations which are far from harmonic. 

Due to the low value of the Hubble constant relative to $\bar{m}_\phi$
and hence the little effect it has on the amplitude of the first
oscillations, in the top right panel in Fig. (\ref{fig:evol}) we can
clearly distinguish the two different parts of each oscillation: the
peak, corresponding to more harmonic-like motion around the minimum,
Eq. (\ref{solnminh}), and the non-oscillating section where the fields
decay as $\sim 1/t^2$ as they approach the initial values. This second
part of the motion is rapidly affected by the expansion of the
Universe as it causes a slight decrease in the amplitude,
and at later stages (top right) it is further suppressed.

\begin{figure}
\scalebox{0.315}{\rotatebox{270}{\includegraphics{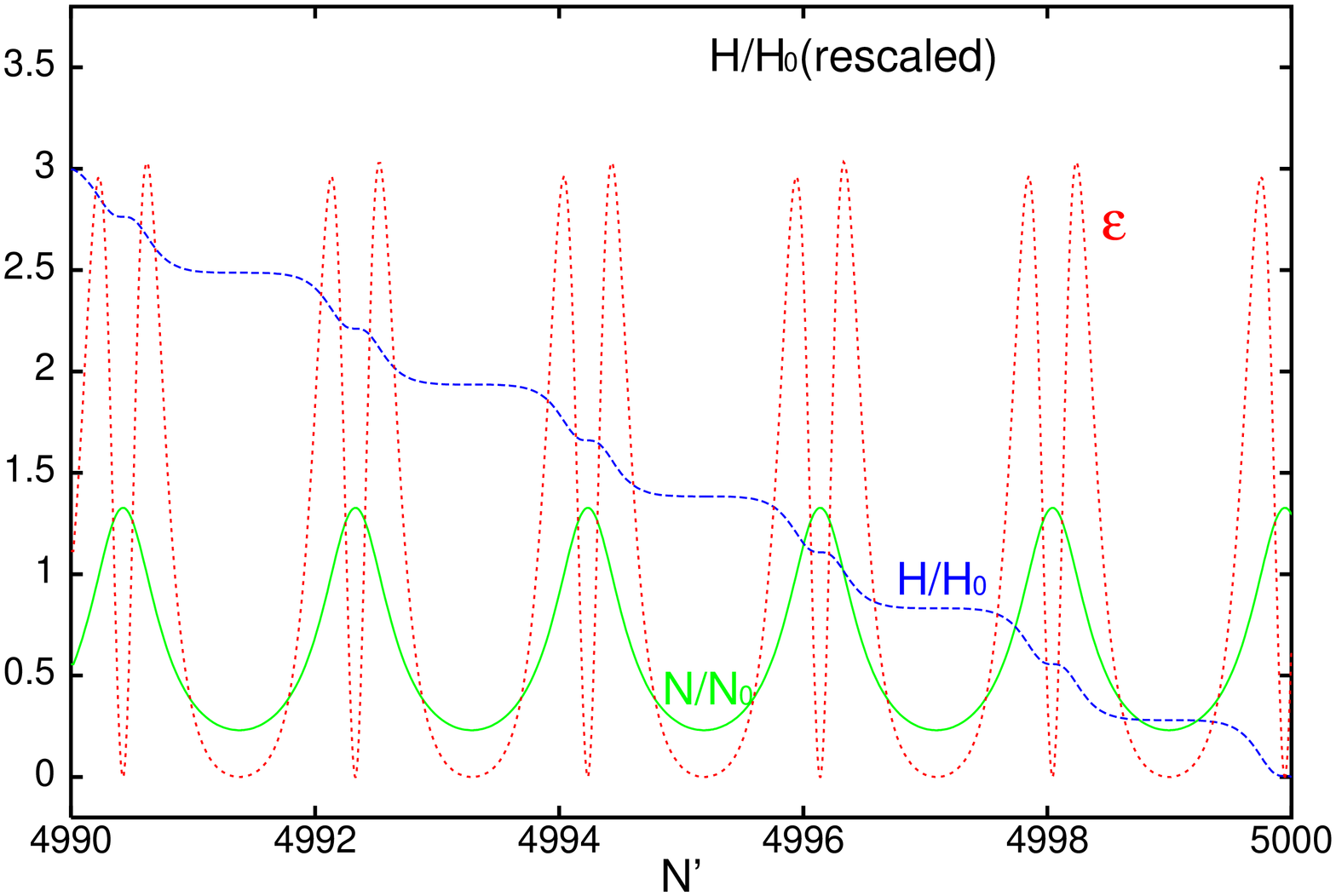}}}
\scalebox{0.315}{\rotatebox{270}{\includegraphics{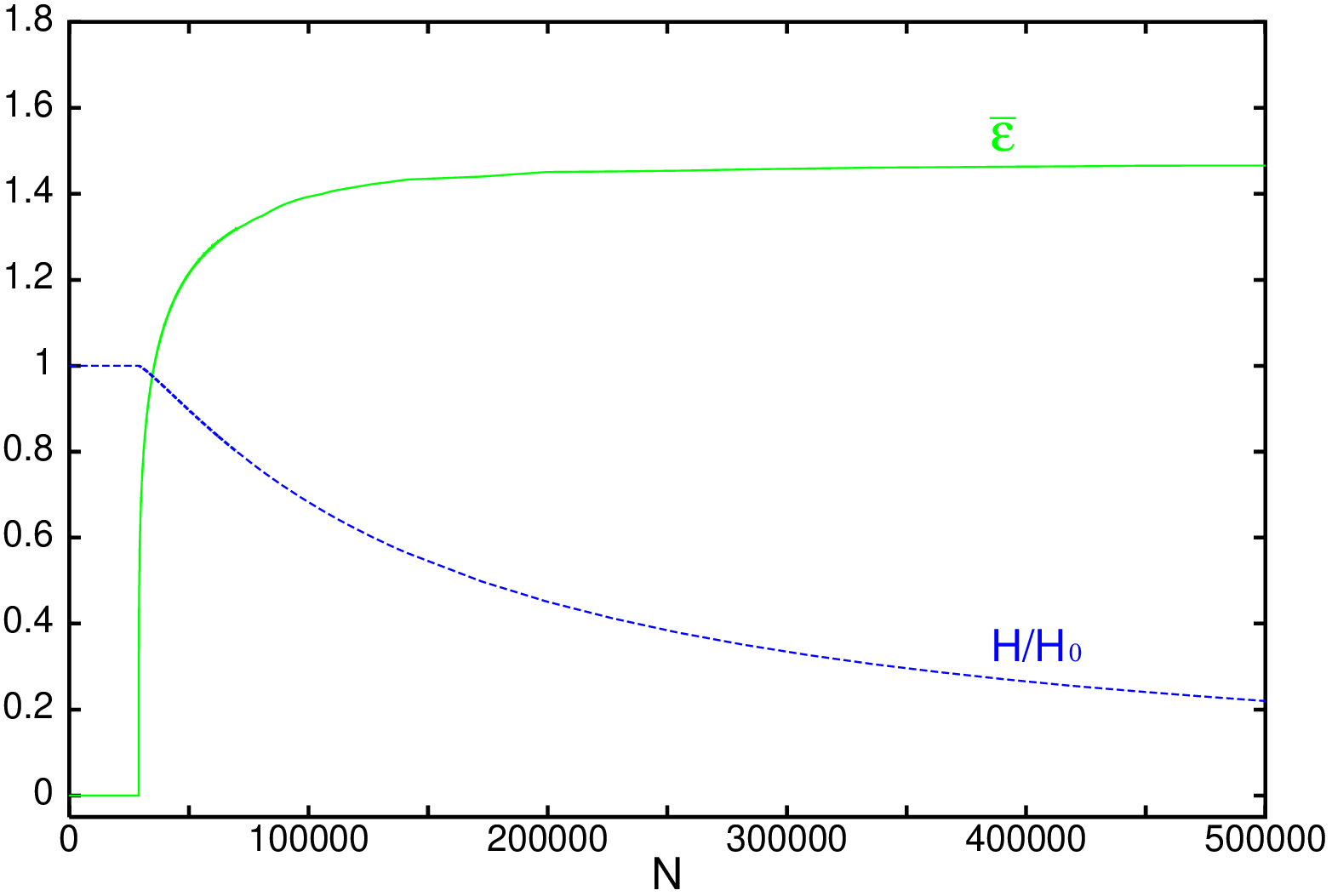}}}
\caption{Behaviour of Hubble constant. Left: Small scale behaviour,
showing oscillations in $\varepsilon$. $H/H_0$ is shown over
the range [0.97814,0.97808], but has been rescaled to make it visible
in the plot. Right: Large scale behaviour - Universe approaching
matter domination ($\bar{\varepsilon}\rightarrow 3/2)$.}
\label{fig:Hub}
\end{figure}

The behaviour of the Hubble constant is shown in Fig.~\ref{fig:Hub}
where we have also plotted the quantity
$\varepsilon=-\dot{H}/{H^2}$. $\varepsilon=0$ corresponds to the
vacuum dominated era of inflation, $\varepsilon=3/2$ to the
reheating phase when matter (specifically the decay products before they
have thermalised) dominates, and $\varepsilon=2$ to the final (after
the reheat temperature has been 
reached) radiation dominated Universe. During the oscillations,
$\varepsilon$ oscillates between the values 0 and 3 and its average
value per oscillation, $\bar{\varepsilon}$ shows the transition between
vacuum and matter dominated states.


\subsection{Quantum effects - Preheating} \label{sec4.2}

Treating $\delta\phi_{\alpha}$ as fluctuations in the presence of the
background fields, $\phi_{\alpha}(t)$ which have the independent
dynamics described in the previous section, we solve the equations
of motion for the quantum fluctuations of all scalar fields in the
theory,
\begin{equation}
\ddot{\psi_k}_\alpha +
\frac{k^2}{a^2}{\psi_k}_\alpha +
\mathcal{M}^2_{\alpha\beta}{\psi_k}_\beta = 0,               \label{psieq1}
\end{equation}
where ${\psi_k}=a^{3/2}{\delta\phi_k}$. Here, the term $\epsilon$ in
Eq. (\ref{psieq}) is negligible since the Hubble constant is so small
relative to $\bar{m}_\phi$ and $\varepsilon \sim O(1)$ (see
Fig. (\ref{fig:Hub})) so $\dot{H}$ is also small in this model. In
fact, particle production will take place in a tiny fraction of a
Hubble time, and we can safely neglect any effect due to the expansion
of the Universe in the following.  Since we are not considering the
quantum fluctuations of the standard Higgs fields, $\delta H_i$, (see
Appendix A), the system above reduces to two sets of two coupled
equations describing the quantum fluctuations of the $N$, $\phi$, the
axion and $X$ fields. The axion and the $X$ field are the
massless/massive pseudoscalars in the model, defined in terms of the
imaginary components of the complex fields,
\begin{eqnarray}
a = N_I\cos\alpha_a - \phi_I\sin\alpha_a\,,& &\nonumber\\
X = N_I\sin\alpha_a + \phi_I\cos\alpha_a\,,& &
\end{eqnarray}
with $\tan \alpha_a = 2 \phi_0/ N_0$. 
The mass squared matrices, $\mathcal{M}^2_{\phi N},
\mathcal{M}^2_{a X}$, are given in Eqs. (\ref{mphiN}) and
(\ref{maX}) in Appendix A.

\begin{figure}
\scalebox{0.315}{\rotatebox{270}{\includegraphics{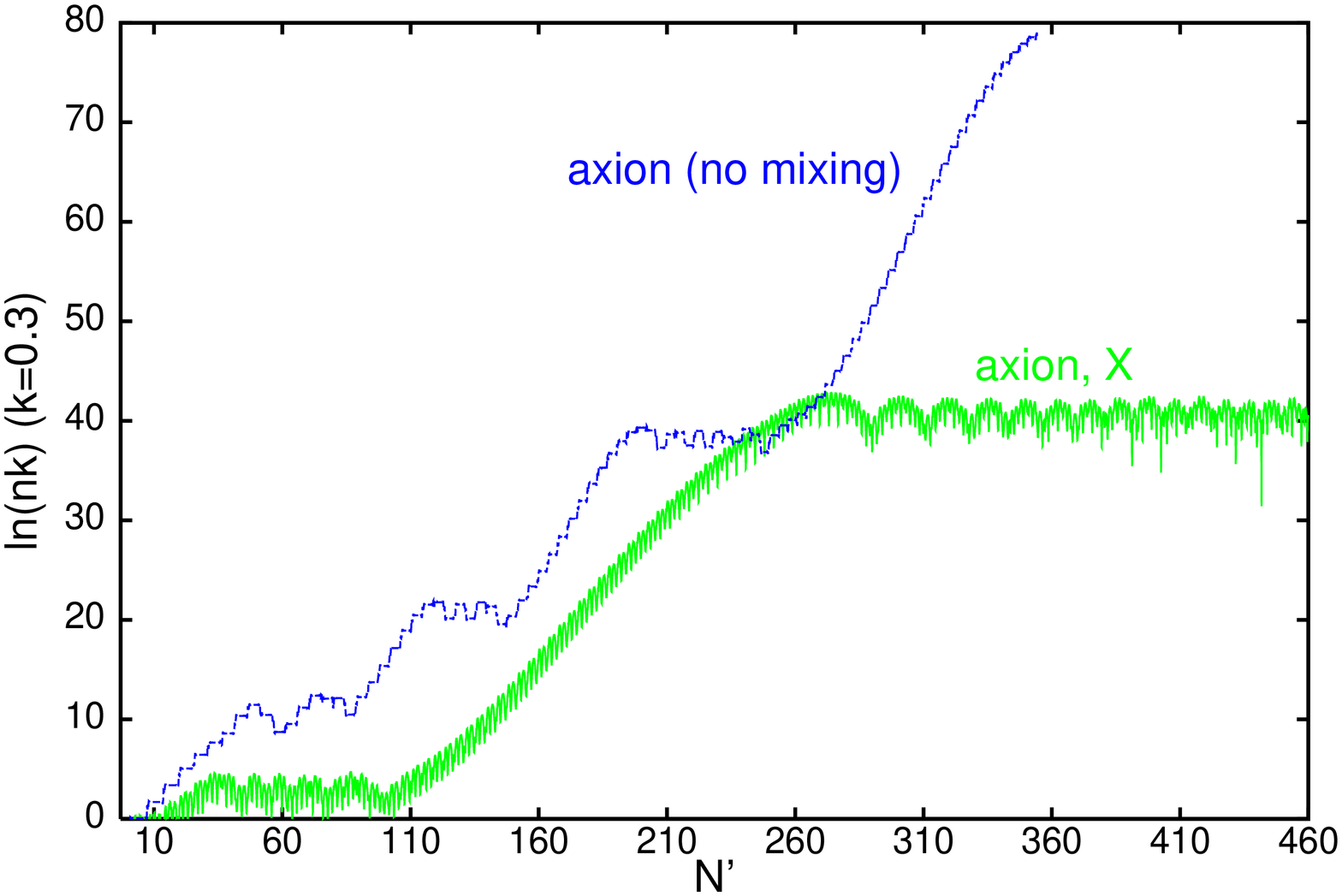}}}
\scalebox{0.315}{\rotatebox{270}{\includegraphics{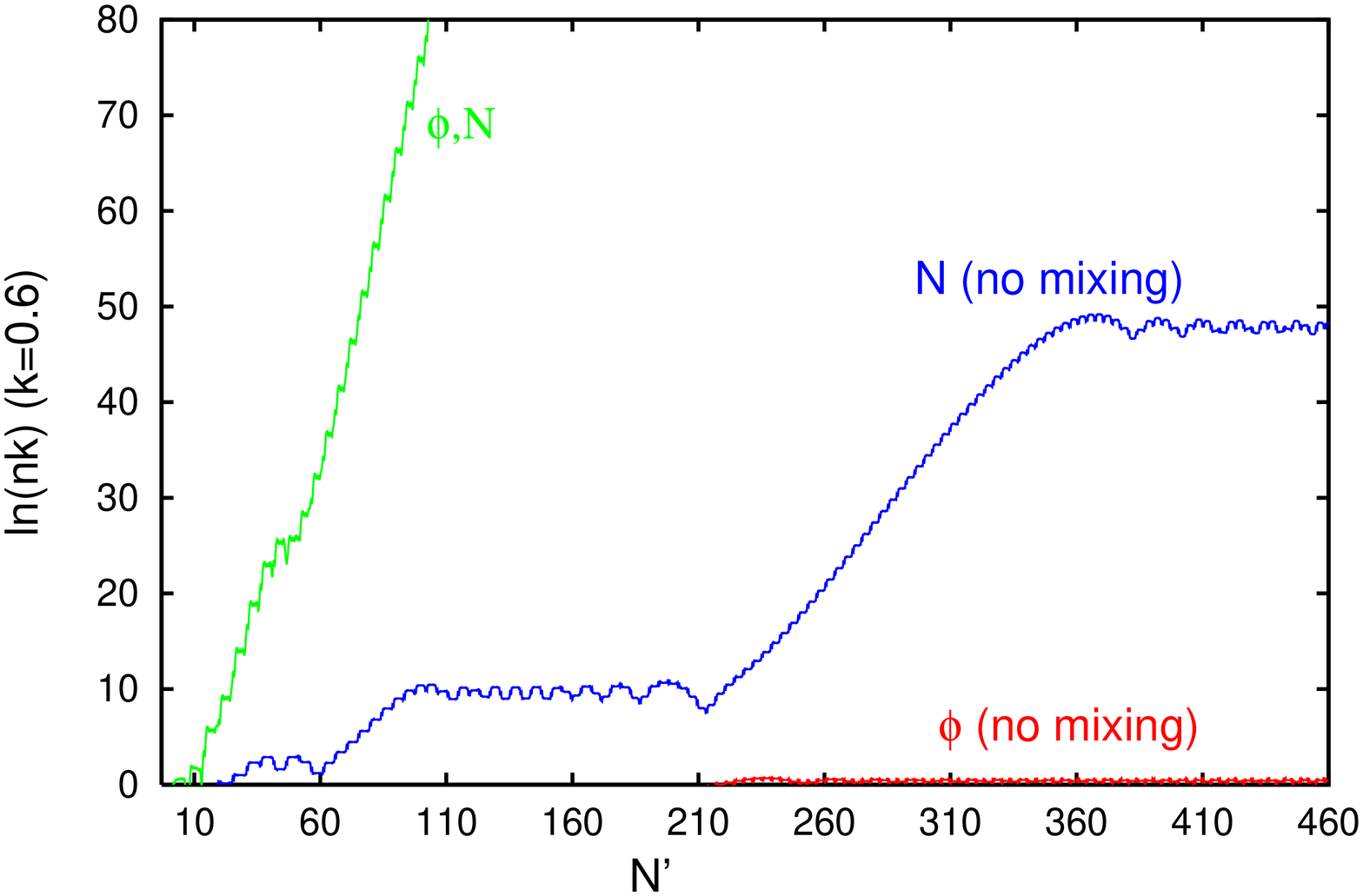}}}
\caption{Explosive production of particles, displayed here by the
exponential increase in the occupation numbers, $n_k$ of axions and
massive $X$ particles for $k=0.3\bar{m}_\phi$ (left); and of
$\phi$ and $N$ particles for $k=0.6\bar{m}_\phi$ (right) with time
(in units of the fundamental period of oscillation of the classical
fields, $N$). The results without including the mixing terms are also
shown for comparison. Without mixing there is no production of
$X$ particles and $\phi$ particle production is
negligible. Including the mixing terms results in virtually identical
numbers of $\phi$ and $N$ particles; and of axions and $X$
particles being produced.}			\label{fig:pprod}
\end{figure}

\begin{figure}
\scalebox{0.315}{\rotatebox{270}{\includegraphics{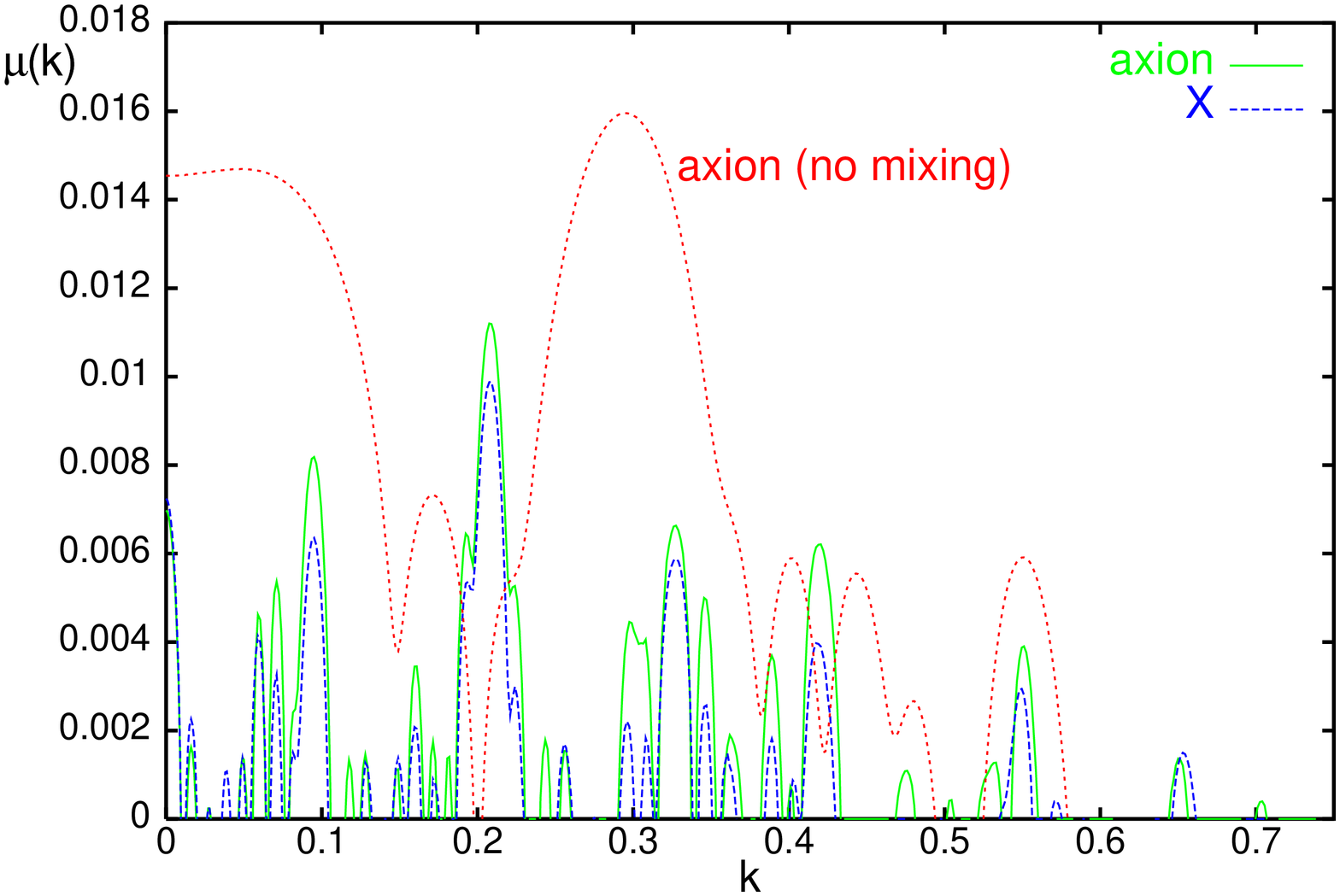}}}
\scalebox{0.315}{\rotatebox{270}{\includegraphics{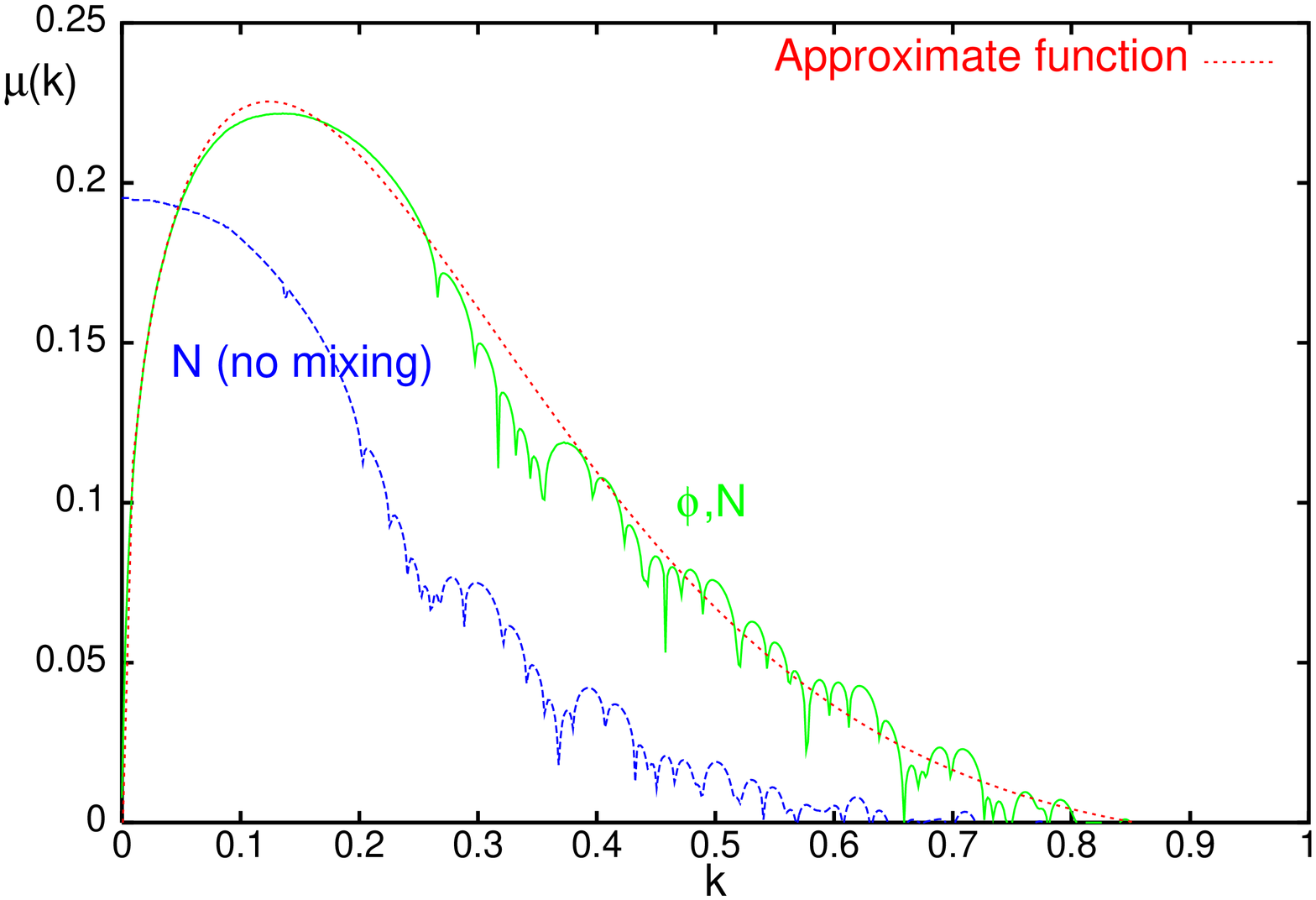}}}
\caption{Typical values of the growth parameter, $\mu_k$ over the
first few ($N'\le 40$) oscillations of the classical fields as a
function of the momentum $k$ in units of $\bar{m}_\phi$.}   \label{fig:muk}
\end{figure}

\begin{figure}
\hfil
\scalebox{0.5}{\rotatebox{270}{\includegraphics{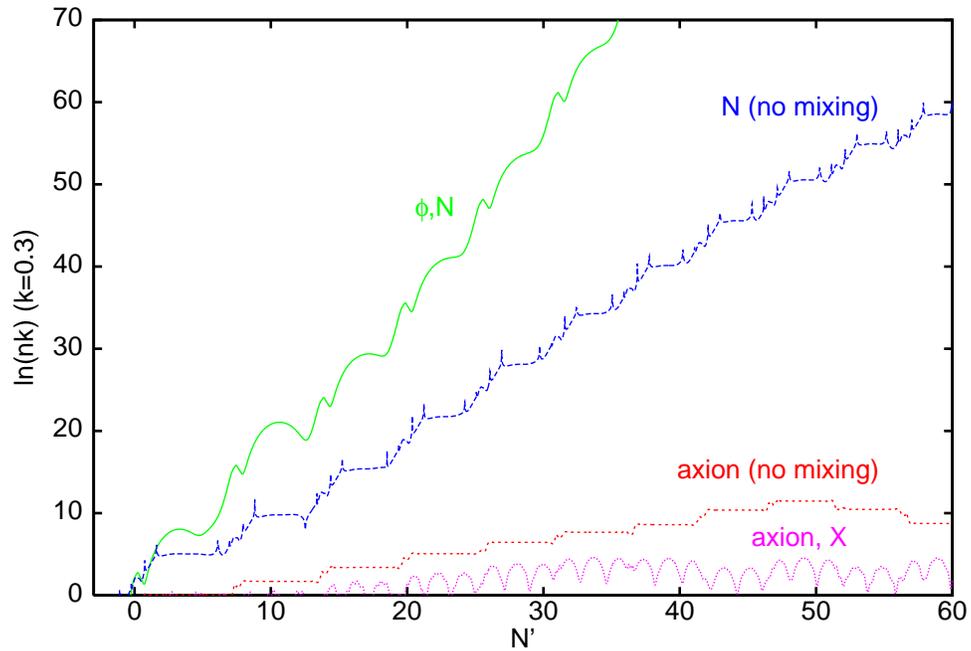}}} \hfil
\caption{Comparison of initial rates of production of all particles for
a typical momentum, $k=0.3\bar{m}_\phi$.}	\label{fig:rates}
\end{figure}

The results obtained by solving the equations numerically for the
parameters given in (\ref{params}) are summarised in Figs. 
(\ref{fig:pprod}-\ref{fig:rates}) where the occupation number $n_k$ is
defined in Eq. (\ref{occno}). In fact particle 
production appears to be extremely efficient. 

We have also plotted for comparison the results obtained treating each
particle separately and solving Eqs. (\ref{psieq1}) as if they were
four uncoupled equations, ie. without the mixing terms ($\alpha\ne
\beta$) in the mass squared matrix $\mathcal{M}^2_{\alpha\beta}$. In
particular in Fig. (\ref{fig:pprod}) we can see that with mixing
there are less axions produced, but now $X$s are also produced at the
same rate, whereas $N$ particle production increases considerably and
now $\phi$ particles are also produced at the same rate. The numerical
results for the $\phi-N$ system fit quite well with the analytical
behaviour obtained in section \ref{pp} using the Mathieu equation .

In the axion-$X$ system, the analysis based on the Mathieu equation
will depend on the details of the model, for instance the value of
$\tan \alpha_a$, and some of the approximations may not be valid, such
as neglecting terms proportional to the cosine squared in the
effective masses. Bearing this in mind, we still can get some
qualitative features studying the equations without the mixing
term. The Mathieu parameters are then given by:
\begin{eqnarray}
q_{a}&\simeq& \frac{4}{9} |\cos^2{\alpha_a}-\sin{2\alpha_a}/\sqrt{2}|
\left(\frac{N_0}{N(t)} \right)^2 
\simeq 0.16 \left(\frac{N_0}{N(t)} \right)^2  , \\
A_{a}(t)&\simeq& \frac{8}{3} \left(\frac{N_0}{N(t)} \right)^2
\frac{k^2}{(a^2\bar{m}_{\phi}^2)},\\ 
q_{X} &\simeq & \frac{4}{9}
|2+\sin^2{\alpha_a}-\sqrt{2}\sin^2{\alpha_a}\tan{\alpha_a}|  
\left(\frac{N_0}{N(t)}
\right)^2 \simeq 0.24 \left(\frac{N_0}{N(t)} \right)^2, \\
A_{X}(t) &\simeq& \frac{8}{3} \left(\frac{N_0}{N(t)} \right)^2 (
\frac{k^2}{a^2\bar{m}_{\phi}^2}+\tan^2{\alpha_a}+1) \simeq 
\frac{8}{3} \left(\frac{N_0}{N(t)} \right)^2
\frac{k^2}{(a^2\bar{m}_{\phi}^2)}  + 56 q_X \,,
\end{eqnarray}
where the RHS numerical factors correspond to $\tan \alpha_a =2$. We
notice that production of $X$ particles would be practically
non-existent, whilst axions may be produced but with $q_a < 0.36$,
which leaves us in the narrow resonance regime. However, these are
only indications of  what could happen if 
the axion and X were not coupled. Due to the mixing term, we  get
that both particles are produced at the same rate, but in this case
the production of axions will be further suppressed. 

\begin{figure}
\hfil
\scalebox{0.5}{\rotatebox{270}{\includegraphics{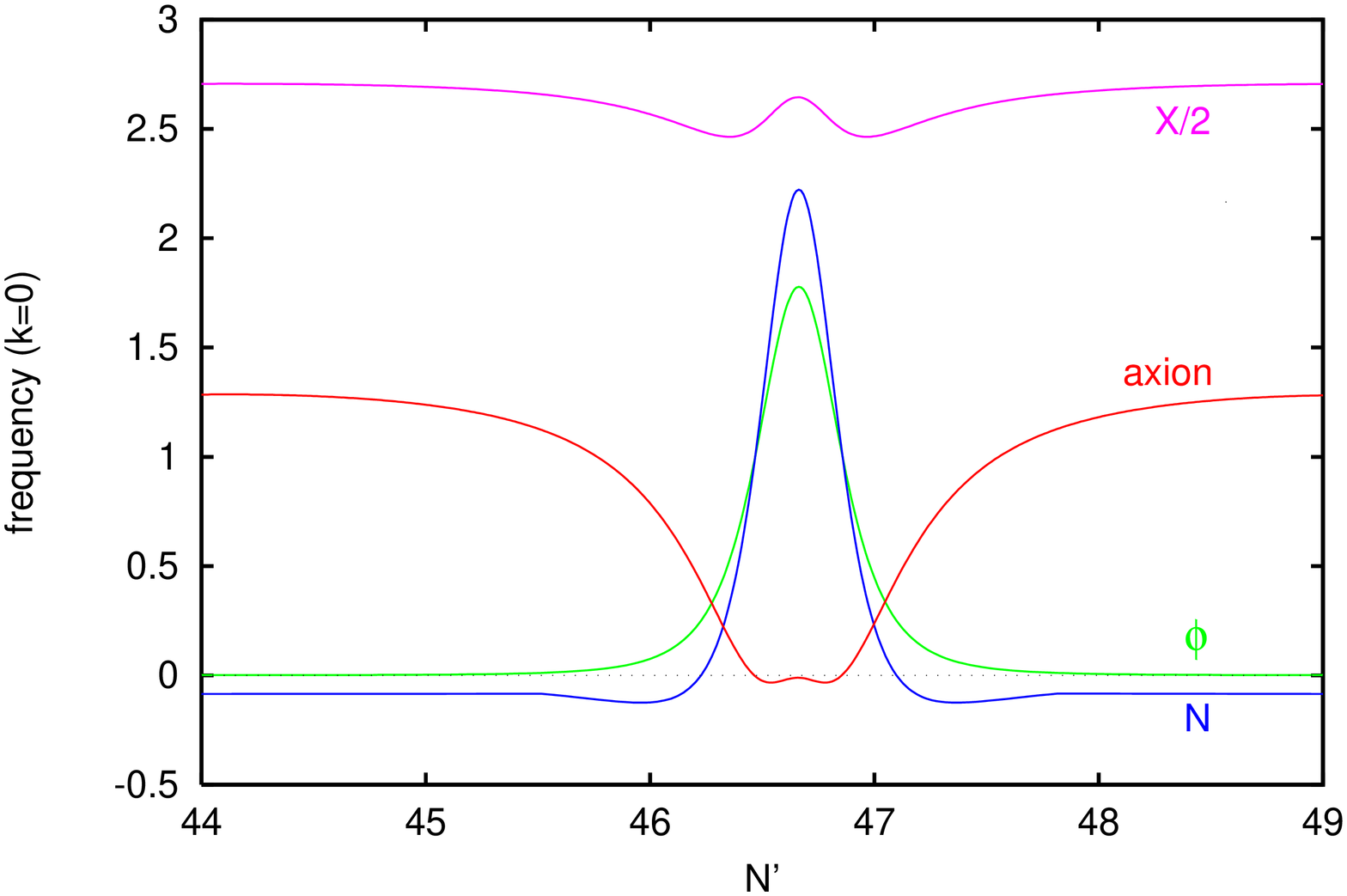}}} \hfil
\caption{The squared frequencies of the quantum fluctuations, $w_k^2$
in units of $\bar{m}_\phi^2$ over one classical field oscillation for
the zero modes ($k=0$) without mixing,
ie. ${w_k}_\alpha^2=\mathcal{M}^2_{\alpha\alpha}$.}  					\label{fig:wks}
\end{figure}

Fig. (\ref{fig:muk}) plots an estimate of the initial rate of growth
of the occupation number of each particle as a function of the
wavenumber $k$, ${\mu_k}_\alpha\sim (4\pi)^{-1}d(\ln
{n_k})_\alpha/dN.$ Without taking into account the mixing terms, the
largest rates correspond to the lowest mode, whilst with the mixing
the effective growth index for $\phi$ and $N$ particles is negligible
for $ k \ll \bar{m}_\phi$, but it has a maximum around $k \approx 0.15
\bar{m}_\phi$. As mentioned in section \ref{comment}, the lower modes
follow the evolution of the homogeneous fields and their production is
suppressed during the oscillations.  For a particular typical value of
$k$, Fig. (\ref{fig:rates}) shows the relative rates of production of
particles. For the beginning oscillations, the classical fields spend
less than 1/6 of the total period of the oscillation around the
minimum, where parametric resonance takes place.  In
Fig. (\ref{fig:rates}), this would correspond to the spikes in the $N$
rate of production with no mixing. Thus, the relatively large values
of $\mu(k)$ for low values of the momentum $k\lsim 0.3 \bar{m}_\phi$
is produced by a simpler mechanism, and is due to the presence of
negative squared frequencies in the evolution equation of the quantum
fluctuations, as can be seen in Fig. (\ref{fig:wks}).  On the other
hand, we want to stress the fact that with mixing the rate of
production for the axion and the $X$ are brought together, but this
now corresponds to a smaller rate of production for the axion. We will
see in the next section that indeed the total number of axions
produced is several orders of magnitude smaller than that of the inflaton
and $N$ particles.

The situation for the pseudoscalars may change
depending on the model. For example, in models where $\phi_0=0$,
($\phi_{c+}= -\phi_{c-}$)
there would be no mixing among the massive and the massless
pseudoscalars and the axion would be the imaginary component of the
$\rm{N}$ field. When particle production begins, the effective squared
mass for the axion would be given by (see Eq. (\ref{mimag}) in
Appendix A), 
\begin{equation}
\mathcal{M}^2_{a}= 2 \kappa^2 N ( N - N_0) \,.
\end{equation}
which again becomes negative whenever $N(t) < N_0$. In this case, we
would find that indeed the production of axions may proceed at the
same rate as that of the inflaton and $N$ particles in a extremely
efficient way.


\subsection{Effects of backreaction. \label{hartree}}

Due to the extremely large numbers of $\phi$ and $N$ particles
produced during the first few oscillations of the background fields,
these particles will quickly give a non-negligible contribution to the
evolution equations. Therefore, instead of the tree-level
equations considered until now, the full dynamics taking into account
the interactions among the particles being produced  should be
included in the evolution equations. In practice this means that the
polarization tensors for the different fields must be
computed beyond the tree-level approximation. A consistent treatment
of all non-linear effects involved can be obtained by using lattice
techniques \cite{ref31,ref3}. However, in order to get some
insight on how backreaction affects the process of preheating, the
Hartree approximation \cite{hartree,hartree2} for the potential should
be good enough.  

Within the Hartree approximation, the main one-loop contributions to
the tadpole and mass terms in the potential are included, not only for
the background fields but for any quantum field. A summary of how to
expand the potential and the new expressions for the effective masses
are given in Appendix B. Including one-loop corrections, the effective
mass of $all$ the fields pick up new non-thermal contributions
proportional to,
\begin{equation}
\langle\delta\phi_\alpha^2\rangle =
\int\frac{k^2dk}{2\pi^2}|{\delta\phi_k}_\alpha|^2.
\label{average} 
\end{equation}
For example, the effective mass for the inflaton field $\phi$ and its
quantum fluctuation will become:
\begin{equation}
\mathcal{M}^2_{\phi \phi}= 2 \kappa^2 (N^2 + \langle\delta N^2\rangle 
+ \cos^2 \alpha_a (\langle \delta a^2 \rangle + t^2 \langle \delta
X^2\rangle + 2 t \langle \delta a \delta X \rangle ) \,, 
\end{equation}
where $t=\tan \alpha_a=2 \phi_0/N_0$ is the mixing angle in the
axion-$X$ sector defined in Appendix A.
The average $\langle \delta \phi_\alpha^2 \rangle$ can be seen as a
closed loop of a particle $\delta \phi_\alpha$. The Hartree
approximation does not cover all possible types of contributions to
the polarization tensors, and the effects of rescattering among the
different species are missed. However, by integrating the coupled system
of equations together with Eq. (\ref{average}), it succeeds in summing up
the infinite set of diagrams of the ``cactus'' or ``daisy'' type. 
The overall effect of the change in the effective masses will be to
shut down the explosive production of particles. On one hand, it will 
increase the frequency of the oscillation of the background fields,
hence decreasing their amplitudes; on the other hand, it will render
the squared frequencies of the quantum fluctuations positive, ending
the process.  

Using the results of the previous section we can estimate the time at
which backreaction becomes important, that is, when $\langle\delta
\phi_\alpha^2\rangle \sim O( \phi_0^2) \sim O (N_0^2)$.  The
expectation values are related to the total number of particles
produced, and the later can be estimated given the growth index
function $\mu(k)$. It is then clear from Fig. (6) that we have to
worry mainly about $\phi$ and $N$ production, and that the total
number of axions and $X$ fields is several orders of magnitude
smaller. This simplifies life, because the function $\mu(k)$ for the
fields $\phi$ and $N$ can be quite nicely approximated by:
\begin{equation}
\mu(k) \approx 0.6\left(\frac{k}{\bar{m}_\phi}\right)^{1/3}exp(-5
(k/\bar{m}_\phi)^{3/2}) \,,  
\end{equation}
valid for the first few oscillations. 
The total number of particles $\phi$ (or $N$) produced at a time
$N'=\bar{m}_\phi t/(2 \pi)$ is then:
\begin{equation}
n_\phi(N') = \frac{1}{4 \pi^2 a^3} \int dk k^2 e^{4 \pi \mu(k) N'}\,,
\end{equation}
and we get $n_\phi \approx O(\phi_0^2)$ when $N' \approx 20 $. Therefore,
backreaction effects will become important just after the third
oscillation of the inflaton field. Soon after that, the results given
in the previous section are modified and quantum fluctuations cease to be
produced at a exponential rate. As a consequence, we will never reach the point
where we have produced too many axions (see Fig. (\ref{fig:pprod})). 

In Fig. (\ref{fig:muk}) it can be seen that only particles with low
momenta, $k < \bar{m}_\phi$, are effectively being produced. Then, the
value $\bar{m}_\phi$ can be used as a cut-off in the momentum
integrals, especially when computing the expectation values
$\langle\delta \phi^2_\alpha\rangle$. In principle, these quantities
are quadratically divergent and have to be renormalised \cite{refren}.
In practice, we avoid this subtle point using the cut-off; the rest of
the couplings and mass parameters appearing in the equations are then
taken to be the renormalised ones.

\begin{figure}[h]
\epsfxsize=12cm
\epsfysize=10cm
\hfil \epsfbox{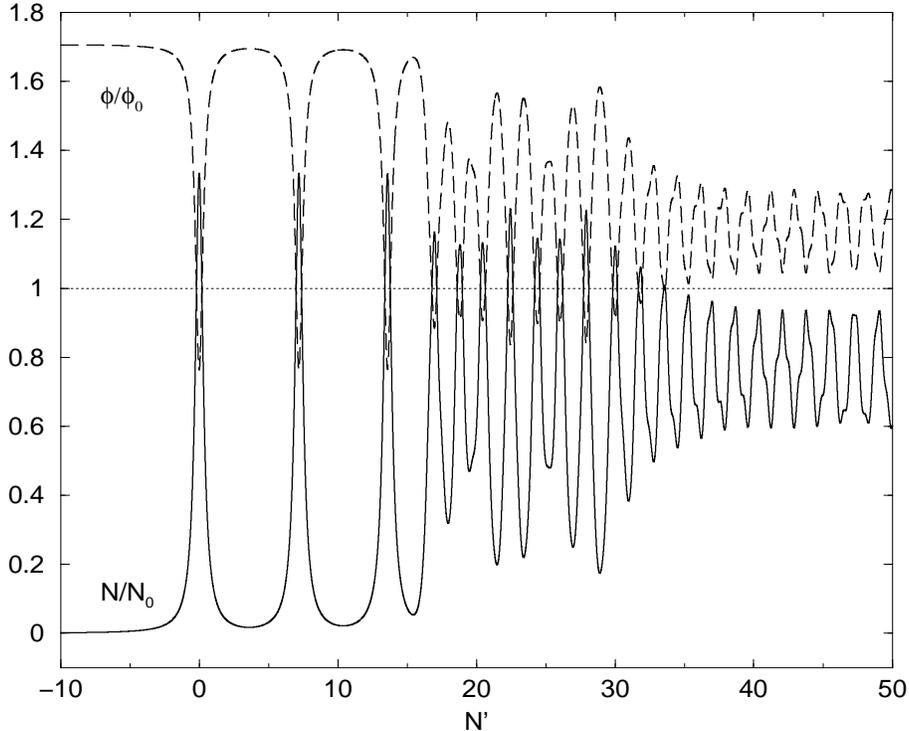} \hfil
\caption{Oscillations of the classical fields with backreaction
effects included.} 
\label{phiNbr}
\end{figure}

The numerical results confirm our estimation. In Fig. (\ref{phiNbr}) it
is shown how the oscillations for the background fields change after
$N' \sim 20$. However, they are still in phase and follow the
trajectory given in Fig. (\ref{fig:traj}).  Recently, it has been
stressed that, in models with symmetry breaking, backreaction of the
quantum fields could restore the original symmetry \cite{symres}, if
they come to dominate the effective mass during the stage of
preheating. This is not the case in our particular model. Backreaction
itself prevents the quantum fluctuations from reaching the stage where
they might restore the symmetry. However, the shape of the potential
changes and when particle production has effectively stopped, the
classical fields end oscillating around a minimum different to
$\phi_0$, $N_0$. We stress that the whole preheating process takes place
within much less than a Hubble time. This means the expansion of the
Universe can be neglected during this time. The decrease in the
amplitude of the oscillating fields in Fig. (\ref{phiNbr}) is purely
due to the presence of the fluctuations.

\begin{figure}[h]
\epsfxsize=12cm
\epsfysize=10cm
\hfil \epsfbox{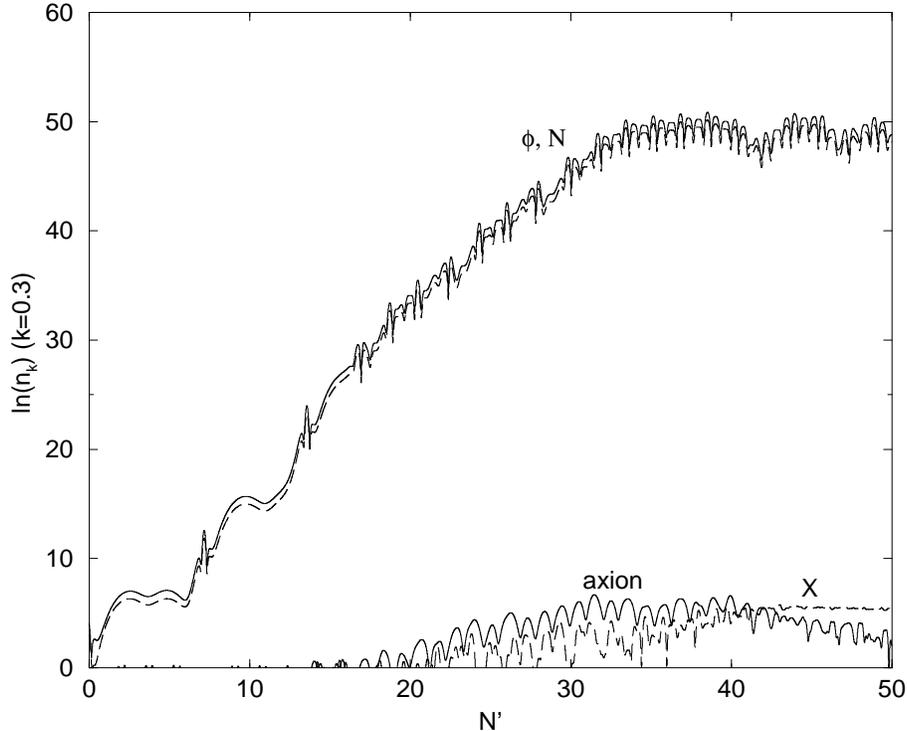} \hfil
\caption{Hartree approximation: Rates of production of all particles
for a typical comoving momentum $k= 0.3 \bar{m}_\phi$.}
\label{nkbr}
\end{figure}

\begin{figure}[h]
\epsfxsize=12cm
\epsfysize=10cm
\hfil \epsfbox{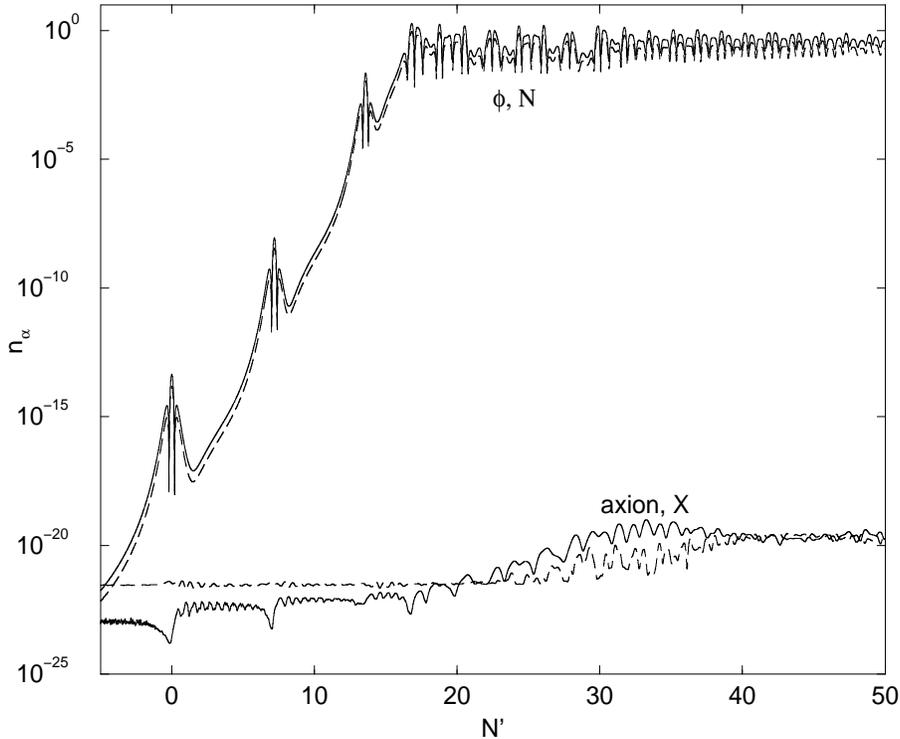} \hfil
\caption{Hartree approximation: total number of particles produced.} 
\label{ntbr}
\end{figure}

In Fig. (\ref{nkbr}) we give the rate of production for all the
different species of particle with a typical momentum $k=0.3
\bar{m}_\phi$. Soon after $N' \sim 20$ they reach a plateau, signaling
the near end of the preheating period. Fig. (\ref{ntbr}) shows the
total number of particles per comoving volume being produced. We end with the ratio of scalars to pseudoscalars being $O(10^{20}$). Preheating
is very efficient in producing $\phi$ and $N$ quanta, and in fact it
is this effect which precludes other particles from being so
efficiently produced.  Backreaction due to $\phi$ and $N$ affects the
evolution of $every$ field (classical and quantum) in the model.

This, however, may not be the end of the story. The Hartree
approximation neglects the effects of rescattering among the particles
produced, but it is possible that these terms could be the same order
of magnitude as the expectation values $\langle\delta
\phi_\alpha^2\rangle$ \cite{ref7}. Lattice simulations \cite{ref3} for
one-field inflation models have shown that the maximum value of
$\langle\delta \phi_\alpha^2\rangle$ is overestimated in the Hartree
approximation when the resonance is very efficient. However, once this
maximum value is reached the resonance ends qualitatively in the same
way as in the Hartree approximation. It is difficult to extrapolate
from these studies to our model. We may have overestimated the total
number of $\phi$ and $N$ produced, and/or it may happen that
rescattering may induce an extra little production of axions and
$X$. But it is unlikely this could reduce the large gap among $n_a$
and $n_\phi$. We can venture that, after preheating, the Universe will
end with the classical fields still displaced from the global minimum,
and fill up with nearly non-relativistic particles, a large fraction
of them being $\phi$ and $N$. A fraction of the energy storage in the
inflaton field at the end of inflation has been transferred to the
particles produced, and this will set the initial conditions for the
subsequent evolution of the system, as given by the standard theory of
reheating; in the Hartree approximation roughly 60\% of the initial
energy density is converted into quantum fluctuations. Moreover, by
the time reheating is complete, any density of stable particles
(i.e. axions) would be diluted by the entropy released in the decay of
the $\phi$ and $N$.

{\section{Conclusion}}

We have analysed in this paper the preheating era which occurs
immediately after the end of inflation in a general SUSY model of
hybrid inflation. From the point of view of the inflationary potential
it is no more than a particular case of the general hybrid model,
Eq. (\ref{Vhyb}). Still, preheating in this particular model is found
to be quite different when studied in more detail. Supersymmetry leads
to equal masses at the global minimum for the real singlets relevant
to inflation ($\phi$ the inflaton and $N$), and therefore to the
presence of a common frequency of oscillation for the classical
fields. That is, the supersymmetric potential allows an approximate
solution with only one mode of oscillation for the classical fields,
the orthogonal one being stationary. This solution becomes exact as
the explicit tiny mass of the inflaton, $m_\phi$, tends to zero. Due to
the inflationary dynamics (slow roll conditions) this mass is indeed
always negligible, independently of the value of any other
parameter. Moreover, the slow roll scenario also fixes the initial
conditions at the beginning of the reheating era such that we begin with
very small initial velocities. Therefore, the fields will follow
almost exactly a straight line trajectory, oscillating with equal
frequencies and proportional amplitudes. Hence, for the purpose of studying
particle production, the SUSY hybrid model can be simplified to a
single field model (the oscillating mode), with the potential given in
Eq. (\ref{Vsingle}).

Since we are dealing with an effective single field oscillating, we
can attempt to convert the evolution equations for the quantum
fluctuations into Mathieu-like equations, whose solutions in term of
stability/instability regions is well known.  We have first studied
production of $\phi$ and $N$ particles, which is completely
independent of the details of the model.  Even in a first
approximation neglecting the effects of backreaction, the evolution
equations are coupled through mixing terms in the effective mass
matrices.  The equations for $\delta \phi$ and $\delta N$ taken
independently (that is, neglecting the mixing term) result in a higher
rate of production of $N$ particles than $\phi$, the latter being
almost completely suppressed. However, the effect of the mixing term
is to bring together the rate of production for both particles, and
overall to become more effective indeed. The fact that we achieve
almost the same number of particles is general to any system of
coupled particles with non negligible mixing. In other words, during
the preheating proccess we can produce particles which in principle
are not expected to exist, if they are coupled to other particle species
which are efficiently produced. The rate of production of the latter
will be increased/decreased depending on the strength of the mixing.

Given that hybrid inflation ends by a phase transition, for some
values of the fields tachyonic masses will be present, for example
that of the $N$ field. As well as an initial production of particles
when the fields are first moving towards the minimum, if the amplitude
of the oscillations is large enough there will be additional
production of particles between oscillations when the classical fields
are in the unstable region of the potential with negative
curvature. Therefore, it is the combination of parametric resonance
and tachyonic mass which causes the production of $\phi$ and $N$
particles to be so efficient.
 
Production of $\phi$ and $N$ fluctuations will rapidly give rise to
non-negligible backreaction effects which will modify the subsequent
evolution of the system. However, other particles may be efficiently
produced in the model which can also contribute to this effect. With
this issue being model dependent, we have proceeded to analyse the
situation in a particle physics based model of inflation, the
$\phi$NMSSM \cite{phiNMSSM}. This model is not only suitable for
inflation, but also solves the strong CP problem by an approximate
Peccei-Quinn symmetry and leads to an invisible axion, which could be
dangerously produced during preheating. However, in this case the
mixing term between the axion and the massive pseudoscalar slightly
suppresses the axion production rate, which turns out to be much
smaller than that of the real scalars. Nevertheless, we obtain the
same general result of comparable rates of production for both
pseudoscalars. Backreaction effects are then dominated by the
production of the real components of the fields, which quite rapidly
succeed in preventing the production of any other particle
species. Therefore, axions in this model are kept safely under control,
thoughtout and after the preheating era.

\begin{center}
{\bf Acknowledgements}
\end{center}
We thank Juan Garc\'{\i}a-Bellido both for motivating this work and for
discussions.

\begin{center}{--------------------------------------------------------------------------}
\end{center}

\appendix
\section{Appendix} 

Here we present some further calculations on the $\phi$NMSSM model,
involving all scalar fields in the theory.

The superpotential,
\[
W = \lambda \rm{N}\rm{H}_1 \rm{H}_2 - \it{{\kappa}}\rm{N}^2\Phi + ...,\nonumber
\]
leads us to the full Higgs part of the tree level scalar potential:
\begin{eqnarray}
V_0(\rm{H}_1,\rm{H}_2,\Phi,\rm{N}) & = & \it{{\kappa}}^{\rm {2}} |\rm{N}|^4 +
(\lambda^2(|\rm{H}_1|^2+|\rm{H}_2|^2) + 4\it{{\kappa}}^{\rm{2}}\rm |\Phi|^2 + 
\it m_N^{\rm 2})|\rm{N}|^2 \nonumber \\ 
  &   & - \it{{\kappa}}A_{\kappa}(\rm\Phi \rm{N}^2+\Phi^*\rm{N}^{*2}) - 2\it{{\kappa}}\lambda(\rm\Phi^*\rm{N}^*\rm{H}_1 \rm{H}_2 + h.c.) + \lambda^2|\rm{H}_1 \rm{H}_2|^2
\nonumber \\ 
  &   & + m_1^2|\rm{H}_1|^2 + \it m_{\rm{2}}^{\rm{2}}|\rm{H}_2|^2 - \lambda {\it{A_\lambda}}(\rm{N}\rm{H}_1 \rm{H}_2 +
\rm{N}^* \rm{H}_1^*\rm{H}_2^*)  \label{fullpotl}\\ 
  &   & + \it m_{\phi}^{\rm{2}}|\rm\Phi|^{\rm{2}} + \it{V_D}(\rm
H_1,H_2), \nonumber
\end{eqnarray}                                           
where $V_D(\rm{H}_1,H_2)$ is the usual $D$-term of the MSSM. The full
spectrum of scalar fields is listed below,
\begin{eqnarray}
\Phi &=& (\phi + i\phi_I)/\sqrt{2},\nonumber\\
{\rm N} &=& (N + iN_I)/\sqrt{2},\nonumber\\
{\rm H_1} &=& \left(\matrix{{\rm H}_1^0\cr {\rm H}_1^-\cr}\right) =
\frac{1}{\sqrt{2}}\left(\matrix{H_1^0 + iH^0_{1\;I}\cr H_1^- +
iH^-_{1\;I}\cr}\right),\label{fields}\\
{\rm H_2} &=& \left(\matrix{{\rm H}_2^+\cr {\rm H}_2^0\cr}\right) =
\frac{1}{\sqrt{2}}\left(\matrix{H_2^+ + iH^+_{2\;I}\cr H_2^0 +
iH^0_{2\;I}\cr}\right),\nonumber		
\end{eqnarray}
with VEVs given by:
\begin{equation}
\langle \Phi \rangle = \phi_0,\;\;
\langle {\rm N} \rangle = N_0,\;\;
\langle {\rm H_1} \rangle = \left(\matrix{\nu_1\cr
0\cr}\right),\;\;
\langle {\rm H_2} \rangle = \left(\matrix{0\cr
\nu_2\cr}\right),
\end{equation}
where the electroweak values, $\nu_1\sim\nu_2\sim 250\, GeV$.

The full $12\times 12$ mass squared matrix for the scalar fields is
defined by,
$\mathcal{M}^2_{\alpha\beta}=\partial^2V/\partial\phi_\alpha\partial\phi_\beta$,
where $\phi_\alpha$ is taken to mean each of the real and
imaginary fields in (\ref{fields}). When the standard Higgs are at
their VEVs, $\mathcal{M}^2$ consists of two lots of $4\times 4$
matrices for the CP-even and CP-odd neutral sectors, and two standard
MSSM $2\times 2$ blocks for the charged mass matrices. All other
entries are zero since the charged sectors do not mix with the neutral
ones and the CP-odd and -even sectors do not mix.

In fact the $4\times 4$ matrices can be further split into two
$2\times 2$ matrices since mixing between the singlets and the
standard neutral Higgs (at least when taken at the VEVs) are
suppressed at the very least by a factor $\mathcal{O}({\kappa})$ with respect
to the other terms, so can be neglected. (Note that in general though,
these terms may be non-negligible and then must be included in 
Eqs. (\ref{queom}), leading to the possibility of parametric
production of Higgs particles and also a change in the production
rates of other particles.) We are then left with standard MSSM
matrices for the CP-even neutral Higgs and also for the CP-odd sector
with the usual massless mode and mixing angle given by,
$\tan\beta=\nu_2/\nu_1$. The mass squared matrix for the real singlet
fields is given by,
\begin{equation}
\mathcal{M}^2_{\phi N}=\left(\matrix{2{\kappa}^2N^2+
m_\phi^2,&4{\kappa}^2N(\phi-\phi_0)\cr
4{\kappa}^2N(\phi-\phi_0),&3{\kappa}^2N^2 +
2{\kappa}^2(\phi-\phi_{c+})(\phi-\phi_{c-})\cr}\right), 
\label{mphiN} 
\end{equation}
where $\phi$ and $N$ denote the real classical fields, the
imaginary parts of which are held at zero. Again, here terms due to
coupling to the standard Higgs are very suppressed with respect to the
other terms so have been neglected.

At the global minimum, (\ref{mphiN}) is diagonal with the masses equal,
\begin{equation}
\bar{m}_{\phi}^2 = \bar{m}_{N}^2 = 2{\kappa}^2N_0^2,		 \label{masses}
\end{equation}
where the bar is used to stress that the masses are at the global
minimum. 

Similarly the CP-odd mass squared matrix for the pseudoscalar
singlet fields, $\phi_I$ and $N_I$ is:
\begin{equation}
\mathcal{M}^2_{\phi_I N_I} =
\left(\matrix{2{\kappa}^2N^2+ m_\phi^2, & 4{\kappa}^2\phi_0N\cr
4{\kappa}^2 \phi_0 N, & {\kappa}^2N^2 +
2{\kappa}^2(\phi+\phi_{c+})(\phi+\phi_{c-})\cr}\right). 
\label{mimag}
\end{equation}
As in the CP-even matrix (\ref{mphiN}), we have neglected the small
couplings to the standard Higgs.  Diagonalising when the fields are at
their VEVs gives us the massless mode (axion, $a$) $\bar{m}_a=0$, and
the massive pseudoscalar, $X$, $\bar{m}_X^2=2{\kappa}^2(4\phi_0^2+N_0^2)$,
with mixing angle, $\alpha_a$ given by,
\begin{equation}
\tan{\alpha_a}=2\phi_0/N_0,					\label{mixang}
\end{equation}
with the axion and it's massive counterpart defined in terms of the
imaginary components, $\phi_I,N_I$, by:
\begin{eqnarray}
a = N_I\cos\alpha_a - \phi_I\sin\alpha_a,& &\nonumber\\
X = N_I\sin\alpha_a + \phi_I\cos\alpha_a.& &
\end{eqnarray}
The mass squared matrix, in the basis $\{a,X\}$, then becomes,
\begin{equation}
\mathcal{M}^2_{a X}=2\cos^2{\alpha_a}\left(\matrix{F&G\cr
G&H\cr}\right),                                           \label{maX}
\end{equation}					
where,
\begin{eqnarray}
& & F=Y_1 + Y_2t^2 - 2Y_3t, \nonumber \\
& & G=(Y_1 - Y_2)t + Y_3(1-t^2), \nonumber \\
& & H=Y_1t^2 + Y_2 + 2Y_3t, \nonumber \\
& & Y_1={\kappa}^2N^2/2+{\kappa}^2(\phi+\phi_{c+})(\phi+\phi_{c-}), \nonumber \\ 
& & Y_2={\kappa}^2N^2+ m_\phi^2/2, \nonumber \\ & & Y_3=2{\kappa}^2\phi_0 N, \nonumber\\
& & t=\tan{\alpha_a}.\nonumber
\end{eqnarray}

\section{Appendix}

With particle production in mind, we now proceed to obtain the
equation of motion in the Hartree approximation.
Written explicitly in terms of the four fields,
\bp, \bn, \ba\ and \bx, the full potential, (\ref{Repotl}) becomes:
\begin{eqnarray}
V&=&V(0) + \frac{1}{4}{\kappa}^2 \bn^4 + {\kappa}^2(\bp -
\phi_{c-})(\bp - \phi_{c+}) \bn^2 + \frac{1}{2} m_\phi^2 \bp^2 \nonumber \\
& &+ \kappa^2 \cos^4 \alpha_a (A \ba^4 + B \ba^3 \bx + C \ba^2 \bx^2 +D
\ba \bx^3 + E \bx^4 )  \nonumber \\
& &+ \kappa^2 \cos^2\alpha_a ( F \ba^2 + G \ba \bx + H \bx^2 ),
\label{V0full} 
\end{eqnarray}
where,
\begin{eqnarray}
& & A = (1/4-t^2),\nonumber \\
& & B = t(2t^2-1),\nonumber \\
& & C = (t^4-(5/2)t^2+1),\nonumber \\
& & D = t(2-t^2),\nonumber \\
& & E = t^2(t^2/4+1),\nonumber
\end{eqnarray}
and $F,G$ and $H$ are as given in Appendix A above with $\phi,\,N$
replaced now by \bp, \bn. The fields are decomposed as in
Eq. (\ref{phidelta}), and the potential is expanded assuming the
factorization:
\begin{eqnarray}
\delta \phi_\alpha \delta \phi_\beta \delta \phi_\gamma & \rightarrow
&  \langle\delta \phi_\alpha \delta \phi_\beta\rangle \delta
\phi_\gamma + 
  \langle\delta \phi_\gamma \delta \phi_\alpha\rangle \delta
\phi_\beta + 
  \langle\delta \phi_\beta \delta \phi_\gamma\rangle \delta
\phi_\alpha
 \nonumber \,, \\
\delta \phi_\alpha \delta \phi_\beta \delta \phi_\gamma \delta
\phi_\eta & \rightarrow &  \langle\delta \phi_\alpha \delta \phi_\beta\rangle \delta 
\phi_\gamma \delta \phi_\eta + {\rm permutations} + {\rm Constant} \,.
\end{eqnarray}
Each term in the factorization can be seen as the contribution from a
one-loop diagram, with the contractions $\langle \cdots\rangle$
accounting for the internal propagator in the loop; only those terms
which correspond to physically allowed diagrams are included in the
expansion.  In this way, one-loop contributions to the tadpole and
masses are included, such that we end with a potential quadratic in
the quantum fluctuations.  Using this potential and the tadpole
constraint $\langle\delta \phi_\alpha\rangle=0$ we obtain the
equations of motion as given in Eq. (2) and Eq. (5).  The first
derivatives $\partial V/\partial \phi_\alpha$, including the
additional terms, are now given by:
\begin{eqnarray}
\left(\frac{\partial V}{\partial\phi}\right)_{Hartree} &=& 
\frac{\partial V_0}{\partial\phi} 
+ 2 \kappa^2 (\phi -\phi_0) \langle\delta N^2\rangle + 4 \kappa^2 N
\langle\delta N \delta \phi\rangle \nonumber \\
& & +2 \kappa^2 (\phi+\phi_0) \cos^2\alpha_a(\langle\delta a^2\rangle
+2 t \langle\delta a \delta X\rangle + t^2 \langle \delta X^2\rangle), \\
\left(\frac{\partial V}{\partial N}\right)_{Hartree} &=& 
\frac{\partial V_0}{\partial N} 
+ \kappa^2 N ( 3 \langle \delta N^2 \rangle + 2 \langle \delta
\phi^2\rangle )    
+ 4 \kappa^2 ( \phi - \phi_0) \langle\delta N \delta \phi\rangle \nonumber \\  
& &+ \kappa^2 \cos^2\alpha_a (F'\langle\delta a^2\rangle +
G'\langle\delta a\delta X\rangle + 
H'\langle\delta X^2\rangle ), 
\label{dV0full}
\end{eqnarray}
with $\partial V_0/ \partial \phi$ and $\partial V_0/\partial N$ as 
given in Eqs. (\ref{derivs}), and we have defined,
\begin{eqnarray}
& & F' = (2t^2+1)N-4 t \phi_0 , \nonumber \\
& & G' = 4 \phi_0 (1-t^2)- 2 tN ,\nonumber \\
& & H' = (t^2+2) N+4 t \phi_0 .\nonumber
\end{eqnarray}

Eqs. (\ref{mphiN}) and (\ref{maX}) for the mass squared matrix
also obtain new contributions such that:
\begin{eqnarray}
{\mathcal{M}^2_{\phi N}}_{Hartree}& = &\mathcal{M}^2_{\phi N}
+\kappa^2 \left(\matrix{ 2 \langle\delta N^2\rangle & 
4 \langle\delta \phi \delta N \rangle \cr 
4 \langle\delta \phi \delta N \rangle & 
3 \langle\delta N^2\rangle + 2 \langle\delta \phi^2\rangle \cr } \right)
\nonumber \\ 
\nonumber \\
& &
\!\!\!\!\!\!\!\!\!\!\!\!\!\!\!\!
+\kappa^2 \cos^2\alpha_a
\left(\matrix{ 2 (\langle\delta a^2\rangle +2 t \langle\delta a \delta
X\rangle +t^2 \langle\delta X^2\rangle) &  
0\cr
\hspace{-2cm} 0 &
\hspace{-2.4cm}{ (2t^2+1) \langle\delta a^2\rangle-2 t \langle\delta a
\delta X\rangle 
+(t^2+2) \langle\delta X^2\rangle } \cr} \right), 
\nonumber\\  
 \\
{\mathcal{M}^2_{a X}}_{Hartree}& = & \mathcal{M}^2_{a X} 
+ \kappa^2 \cos^2\alpha_a 
\left(\matrix{ (1+2 t^2) \langle\delta N^2\rangle + 2 \langle \delta
\phi^2\rangle &  
 2 t\langle\delta \phi^2\rangle -  t \langle\delta N^2\rangle \cr  
  2t \langle\delta \phi^2\rangle -  t \langle\delta N^2\rangle  &
 (2+ t^2) \langle\delta N^2\rangle + 2t^2 \langle \delta \phi^2\rangle
\cr} \right) 
\nonumber \\  
\nonumber \\
& & 
\!\!\!\!\!\!\!\!\!\!\!\!\!\!\!\!\!\!\!\!\!\!\!\!\!\!\!\!\!\!\!\!\!\!\!\!\!\!
+\kappa^2 \cos^4\alpha_a
\left(\matrix{12 A \langle\delta a^2\rangle+6 B \langle\delta a \delta X\rangle+2 C \langle\delta X^2\rangle &
3 B\langle\delta a^2\rangle+4 C\langle\delta a \delta X\rangle+3 D \langle\delta X^2\rangle \cr
3 B\langle\delta a^2\rangle+4 C\langle\delta a \delta X\rangle+3 D \langle\delta X^2\rangle &
2 C\langle\delta a^2\rangle+6 D\langle\delta a \delta X\rangle+12 E\langle\delta X^2\rangle \cr }\right) 
.\nonumber\\
\end{eqnarray}

In addition, the Friedmann equation for a Universe containing
classical fields and particles now reads:
\begin{equation}
H^2(t) =
\frac{8\pi}{3M_P^2}\left[
\frac{1}{2}\sum_{\alpha}\dot{\phi_{\alpha}}^2+V+\sum_\beta \rho_\beta \right], 
\end{equation}
with $\rho_\beta$ being the energy density of the different particles
produced, i.e.,
\begin{equation}
\rho_\beta=\int\frac{k^2dk}{2\pi^2 a^3} \omega_\beta n_\beta .
\end{equation}

\end{document}